# The Dirac Algebra and its Physical Interpretation

Peter Rowlands* and J. P. Cullerne†

*IQ Group and Department of Physics, University of Liverpool, Oliver Lodge Laboratory, Oxford Street, P.O. Box 147, Liverpool, L69 3BX, UK.  e-mail prowl@hep.ph.liv.ac.uk and prowl@csc.liv.uk

†IQ Group, Department of Computer Science, University of Liverpool, Chadwick Laboratory, Peach Street, Liverpool, L69 72F, UK.

*Abstract.* A version of the Dirac equation is derived from first principles using a combination of quaternions and multivariate 4-vectors. The nilpotent form of the operators used allows us to derive explicit expressions for the wavefunctions of free fermions, vector bosons, scalar bosons, Bose-Einstein condensates, and baryons; annihilation, creation and vacuum operators; the quantum field integrals; and C, P and T transformations; and to suggest new insights into the meaning of supersymmetry and renormalization.

**1 Introduction**

The algebra required for the quantum mechanical equation for the electron, as originally introduced by Dirac, has been discussed by many authors on a formal basis. Our aim, here, is to derive it from a more purely physical starting point in a fundamental symmetry that we believe exists between the four fundamental parameters space, time, mass and charge. This symmetry, we believe, may also be significant in explaining many aspects of the Standard Model of particle physics. An important feature of the symmetry is that it requires a quaternion representation for mass and the three types of charge (electromagnetic, strong and weak), on the same basis as we apply a 4-vector representation of space and time, though the three types of charge must also be separately conserved, unlike the three dimensions of Euclidean space. In applying these two representations simultaneously, we generate a quaternion-multivariate-4-vector algebra, which leads to a mathematical representation of the Dirac wavefunction as a 4-spinor composed of nilpotent terms.

In principle, we show that this is the most complete representation of the Dirac wavefunction available, as its operators are already quantum field operators, eliminating the need for second quantization. We derive explicit versions of the Dirac wavefunctions for free fermions, vector bosons, scalar bosons, Bose-Einstein condensates, and baryons, together with their intrinsic parities; annihilation, creation and vacuum operators; the quantum field integrals; and C, P and T transformations. Standard results, such as spin, helicity, the hydrogen atom, and the Schrödinger approximation, are easily obtained, while the fact that our operators are also supersymmetry operators leads to a completely new insight into the meaning of supersymmetry and renormalisation. At the same time, the relation between our representation and other standard ones is immediately apparent. These results, we believe, arise only because we have started with a formalism which is *physically*, as well as mathematically, true, and leads to a deeper understanding of the physical meaning and origin of the Dirac quantum state.

## 2 The Dirac algebra

Two symmetrical algebras are the basis of the work in this paper. These are the algebras of 4-vectors, with real vector units **i**, **j**, **k** and imaginary scalar $i$; and quaternions, with imaginary vector units $i$, $j$, $k$ and real scalar 1. The imaginary quaternions follow the usual multiplication rules:

$$i^2 = j^2 = k^2 = ijk = -1$$
$$ij = -ji = k$$
$$jk = -kj = i$$
$$ki = -ik = j,$$

while the real vector units will be assumed to follow multivariate multiplication rules, identical to those for Pauli matrices, and parallel to those for quaternion algebra:

$$\mathbf{i}^2 = \mathbf{j}^2 = \mathbf{k}^2 = 1$$
$$\mathbf{ij} = -\mathbf{ji} = i\mathbf{k}$$
$$\mathbf{jk} = -\mathbf{kj} = i\mathbf{i}$$
$$\mathbf{ki} = -\mathbf{ik} = i\mathbf{j}.$$

In effect, this means defining a 'full product' for two vectors **a** and **b** of the form

$$\mathbf{ab} = \mathbf{a}.\mathbf{b} + i\,\mathbf{a} \times \mathbf{b}.$$

This extension of the vector system to the full multivariate form will be significant in accommodating the concept of spin, and is based on the idea that the two algebras used must be completely symmetrical.

It can be readily shown that the algebra of the gamma matrices used in the Dirac equation

$$(\gamma^\mu \partial_\mu + im)\,\psi = 0,$$

is completely isomorphic to a combination of the quaternion and 4-vector algebras, with assignments of the form

$$\begin{array}{ll}
\gamma^0 = -i\mathbf{i} & \gamma^0 = i\mathbf{k} \\
\gamma^1 = \mathbf{i}k & \gamma^1 = \mathbf{i}i \\
\gamma^2 = \mathbf{j}k & \gamma^2 = \mathbf{j}i \\
\gamma^3 = \mathbf{k}k & \gamma^3 = \mathbf{k}i \\
\gamma^5 = ij & \gamma^5 = ij.
\end{array}$$

(or between the two columns)

In general, we will use the first set for the conventional form of the equation, obtaining our altered form by multiplying from the left throughout by $ij$. The cyclic nature of the quaternion operators then allows us to redefine our $\gamma$ terms in the form of the second set, which will be used for the altered form of the equation. We will, however, derive our altered form directly from first principles before relating it to the conventional form.

The complete algebra has 32 parts: 1 real scalar, 1 imaginary scalar, 3 real vectors, 3 imaginary vectors, 3 quaternions, 3 imaginary quaternions, 9 real vector quaternions and 9 imaginary vector quaternions.[1,2,3] The existence of 32 parts suggests that it can be generated from a binomial combination of five 'primitive' (though composite) components, of which the $\gamma$ matrices are a characteristic set, with terms of the opposite sign obtained by reversing the order of multiplication:

$\gamma^0 = i\mathbf{k},\ \gamma^1 = i\mathbf{i},\ \gamma^2 = i\mathbf{j},\ \gamma^3 = i\mathbf{k},\ \gamma^5 = i\mathbf{j}$,

$\gamma^0\gamma^1 = ij\mathbf{i},\ \gamma^0\gamma^2 = ij\mathbf{j},\ \gamma^0\gamma^3 = ij\mathbf{k},\ \gamma^0\gamma^5 = i,\ \gamma^1\gamma^2 = -i\mathbf{k}$,
$\gamma^1\gamma^3 = i\mathbf{j},\ \gamma^1\gamma^5 = i\mathbf{ki},\ \gamma^2\gamma^3 = -i\mathbf{i},\ \gamma^2\gamma^5 = i\mathbf{kj},\ \gamma^3\gamma^5 = i\mathbf{kk}$,

$\gamma^0\gamma^1\gamma^2 = k\mathbf{k},\ \gamma^0\gamma^1\gamma^3 = -k\mathbf{j},\ \gamma^0\gamma^1\gamma^5 = \mathbf{i},\ \gamma^0\gamma^2\gamma^3 = k\mathbf{i},\ \gamma^0\gamma^2\gamma^5 = \mathbf{j}$,
$\gamma^0\gamma^3\gamma^5 = \mathbf{k},\ \gamma^1\gamma^2\gamma^3 = -i\mathbf{i},\ \gamma^1\gamma^2\gamma^5 = j\mathbf{k},\ \gamma^1\gamma^3\gamma^5 = -j\mathbf{j},\ \gamma^2\gamma^3\gamma^5 = j\mathbf{i}$,

$\gamma^0\gamma^1\gamma^2\gamma^3 = \mathbf{j},\ \gamma^0\gamma^1\gamma^2\gamma^5 = -ii\mathbf{k},\ \gamma^0\gamma^1\gamma^3\gamma^5 = ii\mathbf{j},\ \gamma^0\gamma^2\gamma^3\gamma^5 = -ii\mathbf{i},\ \gamma^1\gamma^2\gamma^3\gamma^5 = \mathbf{k}$,

$\gamma^0\gamma^1\gamma^2\gamma^3\gamma^5 = -i$.

Since there is reason to suppose that space, time, mass and charge are the most fundamental set of parameters available to physics,[4,5,6] we might expect that the 32-part algebra containing their units is of fundamental physical significance, and that significant physics will also be contained in the 5-fold structure represented by the binomial 'primitives'.

It is certainly possible, using this algebra, to derive the Dirac equation from the relativistic momentum-energy conservation equation

$$E^2 - p^2 - m^2 = 0 ,$$

by, first factorizing and attaching the exponential term $e^{-i(Et - \mathbf{p}.\mathbf{r})}$, so that

$$(kE + ii\ \mathbf{p} + ij\ m)\ (kE + ii\ \mathbf{p} + ij\ m)\ e^{-i(Et - \mathbf{p}.\mathbf{r})} = 0 ,$$

and then replacing $E$ and $\mathbf{p}$ in the first bracket with the quantum operators, $i\partial / \partial t$ and $-i\nabla$, to give

$$\left(ik\frac{\partial}{\partial t} + i\nabla + ijm\right)(kE + ii\ \mathbf{p} + ij\ m)\ e^{-i(Et - \mathbf{p}.\mathbf{r})} = 0 .$$

This can be written in the form

$$\left(ik\frac{\partial}{\partial t} + i\nabla + ijm\right)\psi = 0 ,$$

where the wavefunction

$$\psi = (kE + ii\ \mathbf{p} + ij\ m)\ e^{-i(Et - \mathbf{p}.\mathbf{r})} ,$$

and the vector elements associated with the second term disappear when we take $\mathbf{p}$ in a preferred direction, as will often be convenient. ($p = \mathbf{1}.\mathbf{p}$ can be replaced by $\mathbf{p}$ if we assume, as is conventional, that only one direction of the vector is well-defined.) With the $m$ term fixed as positive, the equation allows for four solutions

$$\psi_1 = (kE + i\mathbf{i}\ \mathbf{p} + i\mathbf{j}\ m)\ e^{-i(Et - \mathbf{p}\cdot\mathbf{r})}$$
$$\psi_2 = (kE - i\mathbf{i}\ \mathbf{p} + i\mathbf{j}\ m)\ e^{-i(Et + \mathbf{p}\cdot\mathbf{r})}$$
$$\psi_3 = (-kE + i\mathbf{i}\ \mathbf{p} + i\mathbf{j}\ m)\ e^{i(Et + \mathbf{p}\cdot\mathbf{r})}$$
$$\psi_4 = (-kE - i\mathbf{i}\ \mathbf{p} + i\mathbf{j}\ m)\ e^{i(Et - \mathbf{p}\cdot\mathbf{r})}\ .$$

which represent the four possible combinations of $\pm E$ (particle / antiparticle) and $\pm \mathbf{p}$ (spin up / down). It can be shown that these solutions are identical to the four produced by the conventional Dirac spinor:

$$\Psi = \begin{pmatrix} \psi_1 \\ \psi_2 \\ \psi_3 \\ \psi_4 \end{pmatrix}$$

and that the new version of the Dirac equation can be derived from the conventional matrix representation and vice versa.

These are the four solutions as obtained using a common differential operator, which may be written:

$$(kE + i\mathbf{i}\ \mathbf{p} + i\mathbf{j}\ m) = \left(ik\frac{\partial}{\partial t} + i\nabla + i\mathbf{j}m\right),$$

using the operator interpretation of $E$ and $\mathbf{p}$. However, using a set of four different differential operators,

$$\left(ik\frac{\partial}{\partial t} + i\nabla + i\mathbf{j}m\right) \qquad \left(ik\frac{\partial}{\partial t} - i\nabla + i\mathbf{j}m\right)$$

$$\left(-ik\frac{\partial}{\partial t} + i\nabla + i\mathbf{j}m\right) \qquad \left(-ik\frac{\partial}{\partial t} - i\nabla + i\mathbf{j}m\right),$$

which is implicit in the Dirac matrix algebra for the nilpotent wavefunction we have defined (see section 4), we may write the equation in matrix form with the solutions:

$$\psi_1 = (kE + i\mathbf{i}\ \mathbf{p} + i\mathbf{j}\ m)\ e^{-i(Et - \mathbf{p}\cdot\mathbf{r})}$$
$$\psi_2 = (kE - i\mathbf{i}\ \mathbf{p} + i\mathbf{j}\ m)\ e^{-i(Et - \mathbf{p}\cdot\mathbf{r})}$$
$$\psi_3 = (-kE + i\mathbf{i}\ \mathbf{p} + i\mathbf{j}\ m)\ e^{-i(Et - \mathbf{p}\cdot\mathbf{r})}$$
$$\psi_4 = (-kE - i\mathbf{i}\ \mathbf{p} + i\mathbf{j}\ m)\ e^{-i(Et - \mathbf{p}\cdot\mathbf{r})}\ .$$

By using nilpotent wavefunctions, in which the algebra of the differential operator is identical to that of the operator part of the wavefunction, we can effectively transfer the signs of $E$ and $\mathbf{p}$ in the exponential term to the differential matrix, which then reduces for each of the individual solutions to the ordinary differential operator. The choice becomes a matter of convention, but it is the existence of four differential operators, in addition to four solutions, which is responsible for the necessity of writing the equation in terms of the product of a $4 \times 4$ matrix and a 4-component vector.

The four solutions represent the four combinations of particle and antiparticle, and spin up and spin down states. Reversal of the sign of **k**E produces the wavefunction for an antiparticle, while reversal of the sign of **ii**p changes the direction of spin. In this form of the equation, the operator **p**, because it is multivariate and is always effectively multiplied by itself in the eigenvalue produced by the differential operator, can represent the momentum vector **p**, the scalar value of $p$, or the spin term σ.**p**. There is no need to multiply the wavefunction operators by any other term, such as an additional spinor, to convert momentum states to spin states; spin is built into the structure of the multivariate operator **p** (see, for example, section 13). It is also of no significance from a fundamental point of view whether or not we specify a preferred direction, as in $p_x$, $p_y$, or $p_z$. We can, in many instances, derive general expressions for **p** from expressions originally derived for such directional terms; the equations for the general and the particular are, in each case, identical. As in conventional theory, to obtain a scalar probability density $\psi\psi^*$, we need to use a combination of all four solutions.

## 3 C-linear maps and lifts

In our expressions for the Dirac equation and the Dirac wavefunction, **p** is understood to be a multivariate momentum vector with the usual three components. The most usual representation for multivariate vectors is the set of Pauli matrices, $\sigma_o$, $\sigma_x$, $\sigma_y$, $\sigma_z$, which means that we may write the Dirac equation as follows:

$$\left(i\mathbf{k}\frac{\partial}{\partial t} + i\,\vec{\sigma}.\,\vec{\nabla} + i\mathbf{j}m\right)\psi = 0,$$

where $\vec{\sigma}.\,\vec{\nabla}$ is understood now to mean the scalar product, $\sigma_x\partial_x + \sigma_y\partial_y + \sigma_z\partial_z$. This has an equivalent 2 × 2 matrix form:

$$\begin{pmatrix} \mathbf{k}E + i\mathbf{i}\,p_z + i\mathbf{j}\,m & i\mathbf{i}\,(p_x - ip_y) \\ i\mathbf{i}\,(p_x + ip_y) & \mathbf{k}E - i\mathbf{i}\,p_z + i\mathbf{j}\,m \end{pmatrix}\begin{pmatrix}\phi\\ \chi\end{pmatrix} = 0,$$

which leads to the coupled linear differential equations

$$(\mathbf{k}E + i\mathbf{i}\,p_z + i\mathbf{j}\,m)\phi + i\mathbf{i}\,(p_x - ip_y)\chi = 0,$$

$$i\mathbf{i}\,(p_x + ip_y)\phi + (\mathbf{k}E - i\mathbf{i}\,p_z + i\mathbf{j}\,m)\chi = 0.$$

Choosing the momentum to be along the z-axis ($p = p_z$), these differential equations decouple to leave

$$(\mathbf{k}E + i\mathbf{i}\,p + i\mathbf{j}\,m)\phi = 0,$$

$$(\mathbf{k}E - i\mathbf{i}\,p + i\mathbf{j}\,m)\chi = 0,$$

with

$$\phi = \psi_1 = (\mathbf{k}E + i\mathbf{i}\,p + i\mathbf{j}\,m)\begin{pmatrix}1\\0\end{pmatrix}e^{-i(Et - pz)},$$

$$\chi = \psi_2 = (kE - i\mathbf{i}\, p + i\mathbf{j}\, m) \begin{pmatrix} 0 \\ 1 \end{pmatrix} e^{-i(Et + pz)}.$$

These are the positive energy solutions; the quaternion representation allows a simple deduction of the negative energy solutions:

$$\psi_3 = (-kE + i\mathbf{i}\, p + i\mathbf{j}\, m) \begin{pmatrix} 1 \\ 0 \end{pmatrix} e^{i(Et + pz)},$$

$$\psi_4 = (-kE - i\mathbf{i}\, p + i\mathbf{j}\, m) \begin{pmatrix} 0 \\ 1 \end{pmatrix} e^{i(Et - pz)}.$$

To carry out quantum mechanical calculations, we need to define a scalar product for the wavefunctions. However, the wavefunctions in the quaternion / spinor form are effectively a tensor product of two representations.

It is convenient to use a C-linear map $\mathbf{F}: H^C \to C^2$ is a mapping from complex quaternions, $H^C$ ($q_\mu \mathbf{i}_\mu$, $q_\mu \in C$) to $C^2$.

$$\mathbf{F}: H^C \to C^2, \quad \mathbf{Q} \to \mathbf{Q} \begin{pmatrix} 1 \\ 0 \end{pmatrix},$$

where $\mathbf{Q}$ is a complex quaternion, and

$$\mathbf{i}_0 = \begin{pmatrix} 1 & 0 \\ 0 & 1 \end{pmatrix}, \quad \mathbf{i}_1 = \mathbf{i} = \begin{pmatrix} 0 & -i \\ -i & 0 \end{pmatrix}, \quad \mathbf{i}_2 = \mathbf{j} = \begin{pmatrix} 0 & -1 \\ 1 & 0 \end{pmatrix}, \quad \mathbf{i}_3 = \mathbf{k} = \begin{pmatrix} -i & 0 \\ 0 & i \end{pmatrix}.$$

Applying this to $\psi_1$, we obtain

$$\psi_1 = \frac{1}{\sqrt{2}} \begin{pmatrix} -i \\ \kappa + i\varepsilon \end{pmatrix} \otimes \begin{pmatrix} 1 \\ 0 \end{pmatrix},$$

where $\kappa = p / E$, $\varepsilon = m / E$, and $\otimes$ is the tensor product between the two representations. Hence all four solutions may take the form

$$\psi_1 = \frac{1}{\sqrt{2}} \begin{pmatrix} -i \\ 0 \\ \kappa + i\varepsilon \\ 0 \end{pmatrix} \quad \psi_2 = \frac{1}{\sqrt{2}} \begin{pmatrix} 0 \\ -i \\ 0 \\ -\kappa + i\varepsilon \end{pmatrix} \quad \psi_3 = \frac{1}{\sqrt{2}} \begin{pmatrix} i \\ 0 \\ \kappa + i\varepsilon \\ 0 \end{pmatrix} \quad \psi_4 = \frac{1}{\sqrt{2}} \begin{pmatrix} 0 \\ i \\ 0 \\ -\kappa + i\varepsilon \end{pmatrix}.$$

This mapping leads to the conventional Dirac equation for these states, which in the case of $\psi_1$ is

$$\begin{pmatrix} -iE & 0 & -im+p & 0 \\ 0 & -iE & 0 & -im-p \\ im+p & 0 & iE & 0 \\ 0 & im-p & 0 & iE \end{pmatrix} \begin{pmatrix} -i \\ 0 \\ \kappa + i\varepsilon \\ 0 \end{pmatrix} = 0 \,.$$

Taking the four solutions, $\psi_1$, $\psi_2$, $\psi_3$, $\psi_4$, as defined above, we combine their values of

$$\psi^{\dagger}\psi = \psi^*\psi \,,$$

to obtain the probability density

$$4\,(-E^2 - p^2 - m^2) = -8E^2 \,,$$

which becomes 1 on application of the normalising factor $i / \sqrt{2E}$ for each solution. $\psi^*$, of course, defines a row vector with the terms, $\psi_1^*$, $\psi_2^*$, $\psi_3^*$, $\psi_4^*$, while $\psi$ is the column vector with $\psi_1$, $\psi_2$, $\psi_3$, $\psi_4$.

For a Dirac wavefunction of the form

$$\psi = (k E + i\mathbf{i}\,\mathbf{p} + i\mathbf{j}\,m)\,e^{-i(Et - \mathbf{p}\cdot\mathbf{r})}$$

we have also defined an Hermitian conjugate,

$$\psi^{\dagger} = \gamma^0 \psi \gamma^0 = \psi^* = (k E - i\mathbf{i}\,\mathbf{p} - i\mathbf{j}\,m)\,e^{i(Et - \mathbf{p}\cdot\mathbf{r})}$$

and an adjoint wavefunction,

$$\overline{\psi} = \psi^{\dagger}\gamma^0 = \psi^{\dagger}\,i\mathbf{k} = (-iE - \mathbf{j}\mathbf{p} + \mathbf{i}\,m)\,e^{i(Et - \mathbf{p}\cdot\mathbf{r})} \,,$$

though it may be equally convenient to replace this with the same function multiplied from the left by $i\mathbf{k}$, so that

$$\overline{\psi} = (k E + i\mathbf{i}\,\mathbf{p} + i\mathbf{j}\,m)\,e^{i(Et - \mathbf{p}\cdot\mathbf{r})} \,,$$

as this function equally satisfies the adjoint equation

$$\left( \frac{\partial \overline{\psi}}{\partial t}\,i\mathbf{k} + \nabla\overline{\psi}\mathbf{i} - \overline{\psi}\mathbf{i}\mathbf{j}m \right) = 0 \,,$$

and is only ever used as a multiplier from the left.

From this, we may derive the bilinear covariants, and hence the current density terms (using the four solution sum and the normalising factor):

$$\begin{aligned} \overline{\psi}\gamma^0\psi &= \overline{\psi}\,i\mathbf{k}\,\psi = i\mathbf{k}\,E^2/E^2 = i\mathbf{k} \\ \overline{\psi}\gamma^1\psi &= \overline{\psi}\,i\mathbf{i}\,\psi = -i\mathbf{i}\,p^2/E^2 \\ \overline{\psi}\gamma^2\psi &= \overline{\psi}\,i\mathbf{j}\,\psi = -i\mathbf{j}\,p^2/E^2 \\ \overline{\psi}\gamma^3\psi &= \overline{\psi}\,i\mathbf{k}\,\psi = -i\mathbf{k}\,p^2/E^2 \\ \overline{\psi}\gamma^5\psi &= \overline{\psi}\,i\mathbf{j}\,\psi = -i\mathbf{j}\,m^2/E^2 \,. \end{aligned}$$

The first four quantities (the scalar and vector terms) are the components of the current-probability density 4-vector, $J^\mu = \bar{\psi}\gamma^\mu\psi$, where $\partial^\mu J^\mu = 0$ in the absence of external fields. Using the original form of the adjoint wavefunction, the scalar, vector and pseudoscalar terms become:

$$\bar{\psi}\gamma^0\psi = 1$$
$$\bar{\psi}\gamma^1\psi = -i\mathbf{j}\mathbf{i}\, p^2/E^2$$
$$\bar{\psi}\gamma^2\psi = -i\mathbf{j}\mathbf{j}\, p^2/E^2$$
$$\bar{\psi}\gamma^3\psi = -i\mathbf{j}\mathbf{k}\, p^2/E^2$$
$$\bar{\psi}\gamma^5\psi = i\, m^2/E^2 ,$$

while the pseudovector terms take the values:

$$\bar{\psi}\gamma^5\gamma^0\psi = -\mathbf{j}\, p^2/E^2$$
$$\bar{\psi}\gamma^5\gamma^1\psi = -\mathbf{i}$$
$$\bar{\psi}\gamma^5\gamma^2\psi = -\mathbf{j}$$
$$\bar{\psi}\gamma^5\gamma^3\psi = -\mathbf{k} ,$$

and the tensor terms are zeroed.

We may also use the bilinear covariants to construct the zero Dirac Lagrangian, which results when the equations of motion are obeyed:

$$\mathcal{L} = \bar{\psi}(i\gamma^\mu\partial_\mu - m)\psi = 0 .$$

**4 The quaternion form derived from a matrix representation**

We have shown, in the previous section, that our quaternion nilpotent algebra can be used to generate the standard version of the Dirac spinors. The reverse process is also possible, but requires a more subtle argument which suggests why exactly four solutions are required (only four are, of course, *possible* if they are to be independent). It also suggests a link may be found with Dirac algebras that structure the four solutions as components of a four-vector or quaternion set.

Here we interpret the matrices as dyadics formed from quaternion (or, alternatively, 4-vector) components, arranged by row and column. Starting with the equation

$$(\alpha \cdot \mathbf{p} + \beta m - E)\, \psi = 0$$

and taking (for convenience, without loss of generality) $p = p_y$, we obtain[8]

$$\alpha_y = \begin{pmatrix} 0 & 0 & 0 & -i \\ 0 & 0 & i & 0 \\ 0 & -i & 0 & 0 \\ i & 0 & 0 & 0 \end{pmatrix}$$

Using[8]

$$\beta = \begin{pmatrix} 0 & 0 & i & 0 \\ 0 & 0 & 0 & i \\ -i & 0 & 0 & 0 \\ 0 & -i & 0 & 0 \end{pmatrix}$$

and applying the unit $4 \times 4$ matrix to $E$, the Dirac equation becomes

$$(\alpha.\mathbf{p} + \beta m - E)\,\psi = \begin{pmatrix} -E & 0 & im & -ip \\ 0 & -E & ip & im \\ -im & -ip & -E & 0 \\ ip & -im & 0 & -E \end{pmatrix} \begin{pmatrix} \psi_1 \\ \psi_2 \\ \psi_3 \\ \psi_4 \end{pmatrix} = 0\,.$$

where the column vector is the usual 4-component spinor, and the terms $E$ and $\mathbf{p}$ are the quantum operators which give the eigenvalues represented by these symbols when applied to the exponential term of the wavefunction. (It will be convenient here to refer to the components of the $4 \times 4$ matrix in terms of these eigenvalues rather than in terms of the operators which produce them.)

We can interpret the rows and columns of this matrix as having either 4-vector or quaternion coefficients. Let us choose the quaternion operators

$$\mathbf{j}, \mathbf{i}, \mathbf{k}, 1$$

as the respective coefficients of the 4 rows. The $4 \times 4$ matrix now becomes a single row bra matrix with the columns:

$$-\mathbf{j}E - i\mathbf{k}m + i\mathbf{p}$$
$$-\mathbf{i}E - i\mathbf{k}p - i m$$
$$i\mathbf{j}m + i\mathbf{i}p - \mathbf{k}E$$
$$-i\mathbf{j}p + i\mathbf{i}m - E\ .$$

If we multiply these terms from the left by the respective *column* coefficients

$$\mathbf{i}, -\mathbf{j}, 1, \mathbf{k},$$

we obtain in each case the expression

$$-\mathbf{k}E + i\mathbf{i}\,\mathbf{p} + i\mathbf{j}\,m\,,$$

which, when multiplied from the right by a wavefunction beginning with this term, gives a zero product. In order to show that this is equivalent to the Dirac equation in matrix form, we need to show that the four solutions $\psi_1, \psi_2, \psi_3, \psi_4$, multiplied by the appropriate quaternion row coefficients, each result in expressions beginning with this term.

Suppose, therefore, that we have the four solutions, as previously assumed:

$$\psi_1 = (kE + i\mathbf{i}\,\mathbf{p} + i\mathbf{j}\,m)\,e^{-i(Et - \mathbf{p}.\mathbf{r})}$$
$$\psi_2 = (kE - i\mathbf{i}\,\mathbf{p} + i\mathbf{j}\,m)\,e^{-i(Et - \mathbf{p}.\mathbf{r})}$$
$$\psi_3 = (-kE + i\mathbf{i}\,\mathbf{p} + i\mathbf{j}\,m)\,e^{-i(Et - \mathbf{p}.\mathbf{r})}$$
$$\psi_4 = (-kE - i\mathbf{i}\,\mathbf{p} + i\mathbf{j}\,m)\,e^{-i(Et - \mathbf{p}.\mathbf{r})}\;.$$

We can show that the terms

$$k\,\psi_1\,k$$
$$-j\,\psi_2\,j$$
$$1\,\psi_3\,1$$
$$i\,\psi_4\,i$$

each produce the expression

$$(-kE + i\mathbf{i}\,\mathbf{p} + i\mathbf{j}\,m)\,e^{-i(Et - \mathbf{p}.\mathbf{r})}\;.$$

So, multiplying each of the terms

$$k\,\psi_1$$
$$-j\,\psi_2$$
$$1\,\psi_3$$
$$i\,\psi_4$$

from the left by $(-kE + i\mathbf{i}\,\mathbf{p} + i\mathbf{j}\,m)$ results in a zero product.

For convenience, we may rearrange these to give a ket matrix of the form:

$$\begin{pmatrix} \psi_1 \\ \psi_2 \\ \psi_3 \\ \psi_4 \end{pmatrix} = i\,\psi_4 - j\,\psi_2 + 1\,\psi_3 + k\,\psi_1$$

where the row coefficients are identical to the column coefficients of the bra matrix (and even allow the elimination of the − sign before −$j$). The resulting equation is identical to the quaternionic Dirac equation, which we have previously derived by direct means, with the four solutions representing the four possible combinations of ± $E$ and ± $\mathbf{p}$ states.

The derivation demonstrates that the reason for the use of $4 \times 4$ matrices is, in fact, the fundamentally quaternionic nature of the Dirac wavefunction. Ultimately, this is because of the 4-vector space-time used in relativistic equations, which is symmetrical with the quaternion algebra for mass and charge. In our understanding, the use of quaternionic operators to define the weak, strong and electromagnetic charges ($w$, $s$, $e$) maps directly on to the use of the same operators for the terms $E$, $\mathbf{p}$, $m$ in the wavefunction, the existence of these terms as independent units stemming ultimately from the separate identities of the fundamental parameters time, space and mass ($T$, $\mathbf{S}$, $M$). Because quaternion operators define the meaning of the rows and columns in the Dirac matrix, the only way we can map charges $w$, $s$, $e$ on to the $E$, $\mathbf{p}$, $m$ terms via these operators is to have $4 \times 4$ matrices, and hence a 4-dimensional space-time signature in the equation.

It is significant that the application of quaternion operators to $w$, $s$, $e$ and $E$, **p**, $m$ − and, by implication, to the more fundamental parameters time, space and mass (or energy) ($T$, **S**, $M$) − is incomplete, each of these groups of three terms requiring a fourth to complete it. We can identify the respective fourth terms, without difficulty, as mass ($M$), angular momentum ($J$) and charge ($Q$). The $4 \times 4$ Dirac matrix, in effect, *incorporates the fourth term as a zero quantity*. But, while the matrix requires four quantities to which the columns and rows apply, the 4-component spinor allows only four possible solutions from their combination. Interpreting the solutions in terms of the number of relative sign combinations of the component terms allows only *three* of the terms to be nonzero. Significantly, the excluded term in each case is an invariant, the system requiring only one invariant quantity (e.g. $m$ or $M$) to demonstrate the variability of the others ($E$, **p**, or $T$, **S**).

**5 Some aspects of the Dirac algebra**

The power of the current version of the Dirac algebra lies in the fact that it uses nilpotents (or square roots of zero) rather than more conventional forms of the Dirac wavefunction. The two may, however, be easily related. Consider the equation:

$$(iE - i\boldsymbol{k}\,\mathbf{p} - im)(iE - i\boldsymbol{k}\,\mathbf{p} + im) = 0, \qquad (1)$$

which is just one version of the usual energy-mass-momentum equation. The two bracketed terms are clearly different, and so are not square roots of zero or nilpotents. However, if we multiply from the left by $-\boldsymbol{j}$ and from the right by $\boldsymbol{j}$ (in the process changing the Clifford algebra of the operators from $Cl_{4,1}$ to $Cl_{2,3}$), we get:

$$-\boldsymbol{j}(iE - i\boldsymbol{k}\,\mathbf{p} - im)(i\mathrm{E} - i\boldsymbol{k}\,\mathbf{p} + im)\boldsymbol{j} = 0.$$

This now becomes:

$$(\boldsymbol{k}E + i\boldsymbol{i}\,\mathbf{p} + i\boldsymbol{j}\,m)(\boldsymbol{k}E + i\boldsymbol{i}\,\mathbf{p} + i\boldsymbol{j}\,m) = 0. \qquad (2)$$

The two bracketed terms are now identical, and so each is a nilpotent or square root of zero. Equation (1) is, in effect, the usual way of writing the Dirac equation, with the left-hand bracket becoming the differential operator and the right-hand bracket the wavefunction, whereas (2) is just as valid, but much more powerful. In incorporating the $m$ term directly, it effectively goes beyond the conventional Dirac wavefunction towards a quantum field interpretation, though it retains, as we have shown, all the physical interpretation available to the conventional term.

It is important to stress here that what is presented as a purely formal development has a physical, rather than mathematical, origin. It is based on the idea that the four fundamental parameters of physics, space, time, mass and charge, have fully symmetrical properties, one aspect of this being that, while space and time can be represented by a 4-vector, mass and charge have a symmetrical formal representation as a quaternion, the real part of which is represented by mass, and the three imaginary parts by the sources of the electromagnetic, strong and weak interactions ($e$, $s$, $w$).[4,5,6] The development of multivariate approaches to vector algebra by Hestenes *et al.*[7] suggests, in addition, that the vector part of the 4-vector space-time is multivariate or quaternion-like, manifesting its additional properties through spin. It then seems reasonable to suppose that what appears to be the most fundamental equation for describing particle states should be described by an algebra combining *fully symmetrical* 4-vectors and quaternions.

Though there are similarities in some aspects to the quaternion and double quaternion algebras used by some authors,[9-17] the differences are significant in creating a version of the Dirac formalism based on nilpotent operators, in particular the extra element of complexification which this requires. The use of vectors, in addition to quaternions, certainly helps to distinguish the two algebraic systems needed, but they are used here because they are believed to be a true physical representation of the components involved, and, at the end of the paper, they will be required for a physical explanation of the Dirac state. The complexification is physically relevant because it comes from the complexification of the $E$ term (or, equivalently, of time), as it does in classical relativistic physics.

The physical connotations of the quaternions and vectors also explain the choice of symbols used, though some readers may find it occasionally difficult to distinguish the vectors **i**, **j**, **k** from the quaternions *i*, *j*, *k*. In fact, for the most part, the vector aspect is subsumed in the **p** term, and we only need the quaternion symbols, retention of the vectors being optional. In addition, vectors and quaternions can be regarded, in some sense, as 'numbers', rather than as merely complicated abstract mathematical objects. This means that we can write down precise expressions for the wavefunctions which have an immediate physical meaning – a meaning which comes from the fact that they originate in a mathematical representation of space-time and mass-charge which is itself physically meaningful.

It is significant that, in using the mathematical formalism which results from this physical origin, it becomes apparent that what the Dirac equation set out to do in 1928 can be achieved more conveniently by using variations from the particular form used at that time. In principle, the Dirac equation, as used by Dirac, requires wavefunctions to be ideals, incorporating only a restricted part of the algebra. The above analysis shows, however, that limiting the wavefunction to an ideal limits drastically the physical interpretation that one can derive from the algebra. In other words, a restriction of the wavefunction to *part* of the algebra is a consequence of the fact that the traditional form of the Dirac equation is incomplete – a fact which results in the necessity of second quantization. We will show, however, in this paper, that second quantization is unnecessary. The Dirac operators as used here are already quantum field operators, and it is the multiplication of the wavefunction by *j* or $i\gamma^5$ to create the nilpotent form that makes this possible. It can also be shown that the operators are, in addition, annihilation and creation operators, vacuum operators, and supersymmetry operators, and that they can explain fermion and boson wavefunctions, spin ½, Pauli exclusion, CPT, etc, and even, it would seem, renormalization.

Now, the 'standard' form of the Dirac wavefunction, as we have defined it, is eigenpotent, when the wavefunction is written in the form it has before postmultiplication by *j*, and each of the four terms in the column vector is an idempotent. One of the strengths of the nilpotent version here defined is that, both the column vector and each of its individual terms is a nilpotent, which is the extreme version of both eigenpotent and idempotent. A minimal left ideal is a column vector of 'primitive' idempotents multiplied from the left, which is a subgroup of some larger group. The (*kE* + *ii***p** + *ijm*) form is a column vector, multiplied from the left, composed of idempotents before multiplication by *j*, and nilpotents (an extreme form of idempotent) after multiplication, and it is, in effect, a subgroup of a larger group. It is an extreme case of an ideal.

The five objects represented by the gammas do not include 1 as such because there are no strictly scalar terms in the Dirac equation in its nilpotent form. The more standard form of Dirac has *i* (complex number) or 1 for the mass term but no $\gamma^5$. The nilpotent formulation effectively replaces 1 or *i* with $\gamma^5$. In the pentads from the double quaternion or vector-quaternion set (say **i**, **j**, **k**, *i*, *j*, *k*), each pentad must incorporate one (and only one) triad such as *i***i**, *i***j**, *i***k**, or equivalent.[18] This is important physically because it 'privileges' the vector nature of **p**. The nilpotent Dirac algebra works for **p** as a vector in 3-D, for *p* as a scalar value, or for σ.**p** in a defined direction as required, and the formulation also works for *E* and **p** incorporating field terms, or producing covariant derivative equivalents (with, of course, the corresponding changes made in the values of *E* and **p**, e.g. $E = \partial/\partial t - ie\phi$ and $\mathbf{p} = -i\nabla + e\mathbf{A}$).

## 6 Fermion and Boson Wavefunctions

The vector-quaternion Dirac equation is simpler and more powerful than the conventional representation, and allows a more direct insight into the equation's physical meaning. In the first place, we have a direct expression involving *E*, **p** and *m* for the wavefunction, and, secondly, we can immediately see the explanation for such things as fermionic- and bosonic-type wavefunctions. Since the exponential terms in the wavefunctions multiply as scalars, and can be expressed, in the nilpotent algebra, in a common form for all wavefunctions, we will often find it convenient, here, to discuss the quaternion operators such as (*kE* + *ii* **p** + *ij m*) as the 'wavefunctions' when we are examining the properties of superposed states.

The immediately obvious aspect of fermionic wavefunctions is that they are composed of noncommutative operators, and hence are clearly antisymmetric. Also, multiplying two identical fermion wavefunctions produces zero, as in

$$(kE + ii\ \mathbf{p} + ij\ m)\ (kE + ii\ \mathbf{p} + ij\ m) = 0\ .$$

As nilpotents, or square roots of zero, the wavefunctions have many advantages over conventional representations. Pauli exclusion is immediately obvious because the product of two identical wavefunctions ψψ, each with four solutions, is zero because each term in it is zero. That is,

$$(kE + ii\ \mathbf{p} + ij\ m)\ (kE + ii\ \mathbf{p} + ij\ m) = 0\ ;$$
$$(kE - ii\ \mathbf{p} + ij\ m)\ (kE - ii\ \mathbf{p} + ij\ m) = 0\ ;$$
$$(-kE + ii\ \mathbf{p} + ij\ m)\ (-kE + ii\ \mathbf{p} + ij\ m) = 0\ ;$$
$$(-kE - ii\ \mathbf{p} + ij\ m)\ (-kE - ii\ \mathbf{p} + ij\ m) = 0\ .$$

Reversing the sign of either *kE* or *ii* **p** or both in one of the factors, however, always produces a nonzero scalar product, such as a multiple of $E^2 + p^2 - m^2$, or $-E^2 + p^2 - m^2$, when summed up over the four solutions. Bosons are superposed states of fermion and antifermion, so are a product of terms with opposite signs of *E* in the components. A vector boson (spin = 1) has fermion and antifermion with the same sign of **p**. So, its four component terms are:

$$(kE + ii\ \mathbf{p} + ij\ m)\ (-kE + ii\ \mathbf{p} + ij\ m)\ ;$$
$$(kE - ii\ \mathbf{p} + ij\ m)\ (-kE - ii\ \mathbf{p} + ij\ m)\ ;$$
$$(-kE + ii\ \mathbf{p} + ij\ m)\ (kE + ii\ \mathbf{p} + ij\ m)\ ;$$
$$(-kE - ii\ \mathbf{p} + ij\ m)\ (kE - ii\ \mathbf{p} + ij\ m)\ .$$

The sum is a nonzero scalar, 4 ($E^2 + p^2 + m^2$) = 8 $E^2$, before normalization. This is also true if the meson is massless, because the sum of the terms:

$$(k E + i \mathbf{i} \, \mathbf{p}) \, (-k E + i \mathbf{i} \, \mathbf{p}) = 0 \,;$$
$$(k E - i \mathbf{i} \, \mathbf{p}) \, (-k E - i \mathbf{i} \, \mathbf{p}) = 0 \,;$$
$$(-k E + i \mathbf{i} \, \mathbf{p}) \, (k E + i \mathbf{i} \, \mathbf{p}) = 0 \,;$$
$$(-k E - i \mathbf{i} \, \mathbf{p}) \, (k E - i \mathbf{i} \, \mathbf{p}) = 0$$

is still nonzero, 4 ($E^2 + p^2$) = 8 $E^2$, before normalization. This means that vector bosons can be massless (e.g. gluons, photons). For a scalar boson (spin = 0) the $p$ terms of fermion and antifermion have opposite signs. Hence the product becomes the sum of:

$$(k E + i \mathbf{i} \, \mathbf{p} + i \mathbf{j} \, m) \, (-k E - i \mathbf{i} \, \mathbf{p} + i \mathbf{j} \, m) \,;$$
$$(k E - i \mathbf{i} \, \mathbf{p} + i \mathbf{j} \, m) \, (-k E + i \mathbf{i} \, \mathbf{p} + i \mathbf{j} \, m) \,;$$
$$(-k E + i \mathbf{i} \, \mathbf{p} + i \mathbf{j} \, m) \, (k E - i \mathbf{i} \, \mathbf{p} + i \mathbf{j} \, m) \,;$$
$$(-k E - i \mathbf{i} \, \mathbf{p} + i \mathbf{j} \, m) \, (k E + i \mathbf{i} \, \mathbf{p} + i \mathbf{j} \, m) \,.$$

This is, again, a scalar, 4 ($E^2 - p^2 + m^2$) = 8 $m^2$, before normalization. However, if we now make the mass zero, we then obtain the sum of:

$$(k E + i \mathbf{i} \, \mathbf{p}) \, (-k E - i \mathbf{i} \, \mathbf{p}) \,;$$
$$(k E - i \mathbf{i} \, \mathbf{p}) \, (-k E + i \mathbf{i} \, \mathbf{p}) \,;$$
$$(-k E + i \mathbf{i} \, \mathbf{p}) \, (k E - i \mathbf{i} \, \mathbf{p}) \,;$$
$$(-k E - i \mathbf{i} \, \mathbf{p}) \, (k E + i \mathbf{i} \, \mathbf{p}) \,.$$

This is zero for a massless particle, 4 ($E^2 - p^2$), and, hence, no scalar boson can have zero mass. If the Higgs, for example, is a scalar boson, then it must have mass. So must the pion, while the massless gauge particles (gluons and photons) must all have spin 1. A boson with a scalar wavefunction will, in addition, be incapable of description via the Dirac equation, as the eigenvalue of a quaternionic quantum operator can only produce a zero product when multiplied by another quaternionic term.

There is one other possibility. If we produce a composite wavefunction, of which both parts are fermions (or antifermions), but spins are opposite, then we get a product which is the sum of:

$$(k E + i \mathbf{i} \, \mathbf{p} + i \mathbf{j} \, m) \, (k E - i \mathbf{i} \, \mathbf{p} + i \mathbf{j} \, m) \,;$$
$$(k E - i \mathbf{i} \, \mathbf{p} + i \mathbf{j} \, m) \, (k E + i \mathbf{i} \, \mathbf{p} + i \mathbf{j} \, m) \,;$$
$$(-k E + i \mathbf{i} \, \mathbf{p} + i \mathbf{j} \, m) \, (-k E - i \mathbf{i} \, \mathbf{p} + i \mathbf{j} \, m) \,;$$
$$(-k E - i \mathbf{i} \, \mathbf{p} + i \mathbf{j} \, m) \, (-k E + i \mathbf{i} \, \mathbf{p} + i \mathbf{j} \, m) \,.$$

This is a nonzero scalar, 4 ($-E^2 - p^2 + m^2$) = $-8 \, p^2$, before normalization, and so theoretically can exist, and this is what happens with a Bose-Einstein condensation.

Baryon wavefunctions may also be constructed from the nilpotents. The standard QCD representation is the antisymmetric colour singlet of *SU*(3):

$$\psi \sim (BGR - BRG + GRB - GBR + RBG - RGB).$$

Here, we use a mapping such as:

|       |                                                   |
|-------|---------------------------------------------------|
| BGR   | $(kE + ij\, m)\ (kE + ij\, m)\ (kE + ii\, \mathbf{p} + ij\, m)$ |
| $-$ BRG | $(kE + ij\, m)\ (kE - ii\, \mathbf{p} + ij\, m)\ (kE + ij\, m)$ |
| GRB   | $(kE + ij\, m)\ (kE + ii\, \mathbf{p} + ij\, m)\ (kE + ij\, m)$ |
| $-$ GBR | $(kE + ij\, m)\ (kE + ij\, m)\ (kE - ii\, \mathbf{p} + ij\, m)$ |
| RBG   | $(kE + ii\, \mathbf{p} + ij\, m)\ (kE + ij\, m)\ (kE + ij\, m)$ |
| $-$ RGB | $(kE - ii\, \mathbf{p} + ij\, m)\ (kE + ij\, m)\ (kE + ij\, m)$ , |

with each term equivalent to $-p^2(kE + ii\, \mathbf{p} + ij\, m)$ or $-p^2(kE - ii\, \mathbf{p} + ij\, m)$.

It is noticeable here that there are three cyclic and three anticyclic combinations, just as in the conventional representation. Elsewhere,[19,20] we have represented the colour exchange process, by the exchange of the **p** term, as in conventional gluon theory. This fits in with **p** and strong charge being isomorphic, both being three-dimensional, but with only one direction well-defined. Because there is only one spin term, it also predicts that the spin is a property of the baryon wavefunction as a whole, not of component quark wavefunctions. This structure is determined solely by the nilpotent nature of the fermion wavefunction. Put in an extra **p** into the brackets missing them, and we immediately reduce to zero.

With the spinor terms included, each of these is represented by a tensor product of three spinors, for example:

$$(kE + ij\, m)\ (kE + ij\, m)\ (kE + ii\, \mathbf{p} + ij\, m) \left(\frac{1}{2}\right) \otimes \left(\frac{1}{2}\right) \otimes \left(\frac{1}{2}\right)$$

where

$$\left(\frac{1}{2}\right) \otimes \left(\frac{1}{2}\right) \otimes \left(\frac{1}{2}\right) = \left(\frac{3}{2}\right) \oplus \left(\frac{1}{2}\right) \oplus \left(\frac{1}{2}\right)$$

So this representation encompasses both spin ½ and spin 3/2 baryon states.

**7 CPT Symmetry**

The quaternionic version of the wavefunction also clarifies the meaning of CPT symmetry. Each of the quaternion operators is responsible for one of the three component symmetries. C, P and T operations on a quaternion wavefunction may be represented as the following operations:

P: $\quad i\,(kE + ii\,\mathbf{p} + ij\,m)\,i = (kE - ii\,\mathbf{p} + ij\,m)$
T: $\quad k\,(kE + ii\,\mathbf{p} + ij\,m)\,k = (-kE + ii\,\mathbf{p} + ij\,m)$
C: $\quad -j\,(kE + ii\,\mathbf{p} + ij\,m)\,j = (-kE - ii\,\mathbf{p} + ij\,m)$ .

From this we may see that:

CP = T: $\quad -j\,(i\,(kE + ii\,\mathbf{p} + ij\,m)\,i)\,j = k\,(kE + ii\,\mathbf{p} + ij\,m)\,k = (-kE + ii\,\mathbf{p} + ij\,m)$
PT = C: $\quad i\,(k\,(kE + ii\,\mathbf{p} + ij\,m)\,k)\,i = -j\,(kE + ii\,\mathbf{p} + ij\,m)\,j = (-kE - ii\,\mathbf{p} + ij\,m)$
TC = P: $\quad k\,(-j\,(kE + ii\,\mathbf{p} + ij\,m)\,j)\,k = i\,(kE + ii\,\mathbf{p} + ij\,m)\,i = (kE - ii\,\mathbf{p} + ij\,m)$

and that TCP $\equiv$ identity, since:

$$k\,(-j\,(i\,(kE + ii\,\mathbf{p} + ij\,m)\,i)\,j)\,k = -kji\,(kE + ii\,\mathbf{p} + ij\,m)\,ijk = (kE + ii\,\mathbf{p} + ij\,m)\ .$$

A violation of charge conjugation, such as happens (at least partially) in the weak interaction, would effectively make the *ij m* term unable to assume the opposite sign, and so would lead to something like

$$\bm{j}\,\psi\,\bm{j} = (\bm{k}\,E + i\bm{i}\,p + i\bm{j}\,m) = \psi.$$

Similar arguments could be applied to parity and time-reversal violation. In such a case, since
$$\bm{j}\,\psi\,\bm{j} = (\bm{k}\,E + i\bm{i}\,p - i\bm{j}\,m) = \psi - 2i\bm{j}\,m,$$

that part of the wavefunction involving the weak interaction would require a term like 2*ij m* to be equivalent to 0. For a neutrino, which only has a weak component of charge, and involves violation of charge conjugation for the entire wavefunction, this would require, in the first instance, a particle with zero mass.

In general, to reduce a term like ± 2*ij m* to 0 requires adding its negative value elsewhere. Weak interactions require such additions for fermions where violation of charge conjugation makes the weak interaction unable to distinguish the weak charge components +*w* and –*w*.

**8 Parities of bosons and baryons**

Defining the parity transformation on $\psi$ as *i* $\psi$ *i*, we can now investigate the intrinsic parities of ground state bosons and baryons. Applying the transformation to a scalar boson, we obtain:

$$i\,(kE + i\bm{i}\,\mathbf{p} + i\bm{j}\,m)\,(-kE - i\bm{i}\,\mathbf{p} + i\bm{j}\,m)\,i\,;$$
$$i\,(kE - i\bm{i}\,\mathbf{p} + i\bm{j}\,m)\,(-kE + i\bm{i}\,\mathbf{p} + i\bm{j}\,m)\,i\,;$$
$$i\,(-kE + i\bm{i}\,\mathbf{p} + i\bm{j}\,m)\,(kE - i\bm{i}\,\mathbf{p} + i\bm{j}\,m)\,i\,;$$
$$i\,(-kE - i\bm{i}\,\mathbf{p} + i\bm{j}\,m)\,(kE + i\bm{i}\,\mathbf{p} + i\bm{j}\,m)\,i.$$

If we take the first term, and use – *i i* = 1, then the parity transformation produces:

$$-i\,(kE + i\bm{i}\,\mathbf{p} + i\bm{j}\,m)\,i\,i\,(-kE - i\bm{i}\,\mathbf{p} + i\bm{j}\,m)\,i\,.$$

Now, we have a parity transformation on each bracket, with an additional – sign. This produces:
$$-\,(kE - i\bm{i}\,\mathbf{p} + i\bm{j}\,m)\,(-kE + i\bm{i}\,\mathbf{p} + i\bm{j}\,m)\,.$$

Applying to each of the terms, we obtain:

$$-\,(kE - i\bm{i}\,\mathbf{p} + i\bm{j}\,m)\,(-kE + i\bm{i}\,\mathbf{p} + i\bm{j}\,m)\,;$$
$$-\,(kE + i\bm{i}\,\mathbf{p} + i\bm{j}\,m)\,(-kE - i\bm{i}\,\mathbf{p} + i\bm{j}\,m)\,;$$
$$-\,(-kE - i\bm{i}\,\mathbf{p} + i\bm{j}\,m)\,(kE + i\bm{i}\,\mathbf{p} + i\bm{j}\,m)\,;$$
$$-\,(-kE + i\bm{i}\,\mathbf{p} + i\bm{j}\,m)\,(kE - i\bm{i}\,\mathbf{p} + i\bm{j}\,m)\,.$$

The total transformed wavefunction *i* $\psi$ *i* thus becomes $-\psi$. The original wavefunction therefore has negative parity.

For the vector meson:

$$- (kE - i\mathbf{i}\,\mathbf{p} + i\mathbf{j}\,m)\,(-kE - i\mathbf{i}\,\mathbf{p} + i\mathbf{j}\,m)\,;$$
$$- (kE + i\mathbf{i}\,\mathbf{p} + i\mathbf{j}\,m)\,(-kE + i\mathbf{i}\,\mathbf{p} + i\mathbf{j}\,m)\,;$$
$$- (-kE - i\mathbf{i}\,\mathbf{p} + i\mathbf{j}\,m)\,(kE - i\mathbf{i}\,\mathbf{p} + i\mathbf{j}\,m)\,;$$
$$- (-kE + i\mathbf{i}\,\mathbf{p} + i\mathbf{j}\,m)\,(kE + i\mathbf{i}\,\mathbf{p} + i\mathbf{j}\,m)\,.$$

Again, $i\psi i$ becomes $-\psi$, and the original wavefunction has negative parity.

Let us try the same operation on a baryon. Taking one of the terms:

$$(kE + i\mathbf{j}\,m)\,(kE + i\mathbf{j}\,m)\,(kE + i\mathbf{i}\,\mathbf{p} + i\mathbf{j}\,m)\,,$$

we apply a parity transformation:

$$i\,(kE + i\mathbf{j}\,m)\,(kE + i\mathbf{j}\,m)\,(kE + i\mathbf{i}\,\mathbf{p} + i\mathbf{j}\,m)\,i\,.$$

This time, we can write it in the form:

$$i\,(kE + i\mathbf{j}\,m)\,i\,i\,(kE + i\mathbf{j}\,m)\,i\,i\,(kE + i\mathbf{i}\,\mathbf{p} + i\mathbf{j}\,m)\,i\,,$$

with no sign change. This term becomes:

$$(kE + i\mathbf{j}\,m)\,(kE + i\mathbf{j}\,m)\,(kE - i\mathbf{i}\,\mathbf{p} + i\mathbf{j}\,m)\,.$$

Taken over all the terms (three with $\mathbf{p}$, and three with $-\mathbf{p}$), then:

$$(kE + i\mathbf{j}\,m)\,(kE + i\mathbf{j}\,m)\,(kE - i\mathbf{i}\,\mathbf{p} + i\mathbf{j}\,m)$$
$$(kE + i\mathbf{j}\,m)\,(kE + i\mathbf{i}\,\mathbf{p} + i\mathbf{j}\,m)\,(kE + i\mathbf{j}\,m)$$
$$(kE + i\mathbf{j}\,m)\,(kE - i\mathbf{i}\,\mathbf{p} + i\mathbf{j}\,m)\,(kE + i\mathbf{j}\,m)$$
$$(kE + i\mathbf{j}\,m)\,(kE + i\mathbf{j}\,m)\,(kE + i\mathbf{i}\,\mathbf{p} + i\mathbf{j}\,m)$$
$$(kE - i\mathbf{i}\,\mathbf{p} + i\mathbf{j}\,m)\,(kE + i\mathbf{j}\,m)\,(kE + i\mathbf{j}\,m)$$
$$(kE + i\mathbf{i}\,\mathbf{p} + i\mathbf{j}\,m)\,(kE + i\mathbf{j}\,m)\,(kE + i\mathbf{j}\,m)\,,$$

and

$$i\,\psi\,i = \psi\,.$$

The baryon wavefunction has positive parity.

These calculations, of course, apply to the ground state values only, because if extra angular momentum terms are added, then extra terms must be supplied to the wavefunctions, the sign of parity reversing for each additional term.

## 9 Annihilation and creation operators

Yet another significant aspect of the quaternion Dirac algebra using nilpotents is that it is already effectively second quantized. We have seen that the quaternion operator $(ik\partial/\partial t + i\nabla + i\mathbf{j}\,m)$ acting on a state

$$(kE + i\mathbf{i}\,\mathbf{p} + i\mathbf{j}\,m)\,e^{-i(Et - \mathbf{p}.\mathbf{r})}$$

leads to the equation:

$$(ik\partial/\partial t + i\nabla + i\mathbf{j}\,m)\,(kE + i\mathbf{i}\,\mathbf{p} + i\mathbf{j}\,m)\,e^{-i(Et - \mathbf{p}.\mathbf{r})} = 0.$$

This operation with operator $\mathcal{D} = (i\boldsymbol{k}\partial/\partial t + i\boldsymbol{\nabla} + i\boldsymbol{j} m)$ may be thought of as a creation operation acting on the single particle fermion state which is already filled. The result is therefore zero. We may obtain the corresponding annihilation operation by finding the Hermitian conjugate of $(i\boldsymbol{k}\partial/\partial t + i\boldsymbol{\nabla} + i\boldsymbol{j} m)$.

We know that:
$$\mathcal{D}\, e^{-i(Et - \mathbf{p}\cdot\mathbf{r})} = (\boldsymbol{k}E + i\boldsymbol{i}\, \mathbf{p} + i\boldsymbol{j}\, m)\, e^{-i(Et - \mathbf{p}\cdot\mathbf{r})} \; . \tag{1}$$

$\mathcal{D}$ is acting as though it were the creation operator $a\dagger$ acting on the group of translations and rotations that we call vacuum. We can therefore write $a\dagger$ as

$$a\dagger = (1/2E)\, (\boldsymbol{k}E + i\boldsymbol{i}\, \mathbf{p} + i\boldsymbol{j}\, m) \; .$$

Now, the Hermitian conjugate of expression (1) is:

$$\mathcal{D}^\dagger\, e^{-i(Et - \mathbf{p}\cdot\mathbf{r})} = (-\boldsymbol{k}E + i\boldsymbol{i}\, \mathbf{p} + i\boldsymbol{j}\, m)\, e^{i(Et - \mathbf{p}\cdot\mathbf{r})} ) \; . \tag{2}$$

Here, because it describes the creation of the anti-particle, $\mathcal{D}^\dagger$ is acting as though it were the annihilation operator $a$ conjugate to $a\dagger$, so that

$$a = (1/2E)\, (-\boldsymbol{k}E + i\boldsymbol{i}\, \mathbf{p} + i\boldsymbol{j}\, m) \; .$$

The operators have been modified by a factor $(1/2E)$ so that, when acting on a state, $a\dagger a$ will reproduce the state up to a scalar multiplicative factor which does not affect the quaternion properties of the state.

It is easy to verify that these two operators have the commutation relations appropriate to fermion annihilation and creation operators:

$$a a\dagger + a\dagger a = 1 \; .$$

We now need to find the vacuum state that $\mathbf{a}^\dagger$ can act upon to lead to the single particle state:
$$(\boldsymbol{k}E + i\boldsymbol{i}\, \mathbf{p} + i\boldsymbol{j}\, m)\, e^{-i(Et - \mathbf{p}\cdot\mathbf{r})} \; .$$

We need only consider the following:

$$a\, (\boldsymbol{k}E + i\boldsymbol{i}\, \mathbf{p} + i\boldsymbol{j}\, m)\, e^{-i(Et - \mathbf{p}\cdot\mathbf{r})} = \mathrm{VAC}$$
$$= 2E\, (1/2E)\, (E - i\boldsymbol{j}\, \mathbf{p} + i\boldsymbol{i}\, m)\, e^{-i(Et - \mathbf{p}\cdot\mathbf{r})} \; .$$

To check this we need only consider

$$a\dagger\, (a\, (\boldsymbol{k}E + i\boldsymbol{i}\, \mathbf{p} + i\boldsymbol{j}\, m)\, e^{-i(Et - \mathbf{p}\cdot\mathbf{r})} = \mathbf{a}^\dagger\, \mathrm{VAC}$$
$$= a\dagger\, (1/2E)\, 2E\, (E - i\boldsymbol{j}\, \mathbf{p} + i\boldsymbol{i}\, m)\, e^{-i(Et - \mathbf{p}\cdot\mathbf{r})}$$
$$= (1/2E)\, 2E\, (\boldsymbol{k}E + i\boldsymbol{i}\, \mathbf{p} + i\boldsymbol{j}\, m)\, e^{-i(Et - \mathbf{p}\cdot\mathbf{r})}$$
$$= (\boldsymbol{k}E + i\boldsymbol{i}\, \mathbf{p} + i\boldsymbol{j}\, m)\, e^{-i(Et - \mathbf{p}\cdot\mathbf{r})} \; .$$

It is easy to verify that the further action of $a$ onto VAC leads to zero.

## 10 The quantum field

Let us define the following:

The creation operator for an electron, spin up:

$$a\dagger(\mathbf{p}) = (\mathbf{k}E + \mathbf{ii}\mathbf{p} + \mathbf{ij}m) / 2E$$

The annihilation operator for an electron, spin up:

$$a(\mathbf{p}) = (-\mathbf{k}E + \mathbf{ii}\mathbf{p} + \mathbf{ij}m) / 2E$$

The annihilation operator for an electron, spin down:

$$a\dagger(-\mathbf{p}) = (\mathbf{k}E - \mathbf{ii}\mathbf{p} + \mathbf{ij}m) / 2E$$

The creation operator for electron, spin up:

$$a(-\mathbf{p}) = (-\mathbf{k}E - \mathbf{ii}\mathbf{p} + \mathbf{ij}m) / 2E$$

The annihilation operator for a positron, spin down:

$$b(\mathbf{p}) = (\mathbf{k}E + \mathbf{ii}\mathbf{p} + \mathbf{ij}m) / 2E$$

The creation operator for a positron, spin down:

$$b\dagger(\mathbf{p}) = (-\mathbf{k}E + \mathbf{ii}\mathbf{p} + \mathbf{ij}m) / 2E$$

The annihilation operator for a positron, spin up:

$$b(-\mathbf{p}) = (\mathbf{k}E - \mathbf{ii}\mathbf{p} + \mathbf{ij}m) / 2E$$

The creation operator for a positron, spin up:

$$b\dagger(-\mathbf{p}) = (-\mathbf{k}E - \mathbf{ii}\mathbf{p} + \mathbf{ij}m) / 2E$$

Now, the anticommutators,

$$\{a\dagger(\mathbf{p}_1), a(\mathbf{p}_2)\} = \delta^3(\mathbf{p}_1 - \mathbf{p}_2)$$

and

$$\{b\dagger(\mathbf{p}_1), a(\mathbf{p}_2)\} = \delta^3(\mathbf{p}_1 - \mathbf{p}_2)$$

are valid for

$$(E_1 - E_2)^2 - (p_1 + p_2)^2 - (m + m)^2 = 0 \ .$$

Writing out the Fourier superpositions of all possible states, we have

$$\psi(x) = \int d^3\mathbf{p}\left\{\left(a(\mathbf{p})\begin{pmatrix}1\\0\end{pmatrix} + a(-\mathbf{p})\begin{pmatrix}0\\1\end{pmatrix}\right)e^{-ipx} + \left(b\dagger(\mathbf{p})\begin{pmatrix}1\\0\end{pmatrix} + b\dagger(-\mathbf{p})\begin{pmatrix}0\\1\end{pmatrix}\right)e^{ipx}\right\}$$

$$\overline{\psi}(x) = \int d^3\mathbf{p}\left\{\left(a\dagger(\mathbf{p})(1\ 0) + a\dagger(-\mathbf{p})(0\ 1)\right)e^{ipx} + \left(b(\mathbf{p})(1\ 0) + b(-\mathbf{p})(0\ 1)\right)e^{-ipx}\right\}$$

Here, because of the explicit expressions used for the creation and annihilation operators, we need only use 1 and 0 states in the spinors.
    Now
$$\psi(x) = |E', \mathbf{p'}\rangle,$$
where
$$a\dagger(\mathbf{p})|0\rangle = |E, \mathbf{p}\rangle$$

$$a\dagger(-\mathbf{p})|0\rangle = |E, -\mathbf{p}\rangle$$

$$b\dagger(\mathbf{p})|0\rangle = |-E, \mathbf{p}\rangle$$

$$b\dagger(-\mathbf{p})|0\rangle = |-E, -\mathbf{p}\rangle$$

$$\psi(x)|E', \mathbf{p'}\rangle = \int d^3\mathbf{p}\left\{\left(a(\mathbf{p})\begin{pmatrix}1\\0\end{pmatrix} + a(-\mathbf{p})\begin{pmatrix}0\\1\end{pmatrix}\right)e^{-ipx}|E', \mathbf{p'}\rangle + \left(b\dagger(\mathbf{p})\begin{pmatrix}1\\0\end{pmatrix} + b\dagger(-\mathbf{p})\begin{pmatrix}0\\1\end{pmatrix}\right)e^{ipx}|E', \mathbf{p'}\rangle\right\}$$

Since
$$a(\mathbf{p})|E', \mathbf{p'}\rangle = a(\mathbf{p})\,a\dagger(\mathbf{p})|0\rangle$$
$$= (\delta^3(\mathbf{p}_1 - \mathbf{p}_2) - a(\mathbf{p})\,a\dagger(\mathbf{p}))|0\rangle$$
$$= |0\rangle,$$

therefore
$$\langle 0|\psi(x)|E', \mathbf{p'}\rangle = \begin{pmatrix}1\\0\end{pmatrix}e^{-ipx}.$$

In principle, therefore, we no longer need to define an explicit process of second quantization. Our nilpotent operators are already second quantized.

**11 Spin**

In addition to such radical reformulations, various standard results can be accommodated into the new formalism by replacing the gamma matrices with quaternion operators. Particularly important is the derivation of fermion spin. In the conventional treatment of spin, we write

$$[\hat{\sigma}, \mathcal{H}] = [\hat{\sigma}, i\gamma_0\gamma\cdot\mathbf{p} + \gamma_0 m] .$$

Also,
$$\hat{\sigma}_l = i\gamma_0\gamma_5\gamma_l , \text{ with } l = 1, 2, 3$$

and
$$i\gamma_0\gamma\cdot\mathbf{p} = i\gamma_0\gamma_1 p_1 + i\gamma_0\gamma_2 p_2 + i\gamma_0\gamma_3 p_3$$

while
$$\gamma_0 = i\mathbf{k}$$
$$\gamma_1 = \mathbf{ii}$$
$$\gamma_2 = \mathbf{ji}$$
$$\gamma_3 = \mathbf{ki}$$
$$\gamma_5 = \mathbf{ij} .$$

So,
$$\hat{\sigma}_1 = -\mathbf{i}$$
$$\hat{\sigma}_2 = -\mathbf{j}$$
$$\hat{\sigma}_3 = -\mathbf{k}$$

or
$$\hat{\sigma} = -\mathbf{1} ,$$

and
$$\gamma = i\mathbf{1} ,$$

where **1** is the unit (spin) vector.

Since $\gamma_0 m = i\mathbf{k}m$ has no vector term and $\hat{\sigma}$ no quaternion, they commute, and we may derive the conventional
$$[\hat{\sigma}, \gamma_0 m] = 0$$

and
$$[\hat{\sigma}, \mathcal{H}] = [\hat{\sigma}, i\gamma_0\gamma\cdot\mathbf{p}] .$$

Now,
$$i\gamma_0\gamma\cdot\mathbf{p} = -\mathbf{j}(\mathbf{i}p_1 + \mathbf{j}p_2 + \mathbf{k}p_3) .$$

So,
$$[\hat{\sigma}, \mathcal{H}] = 2\mathbf{j}(\mathbf{ij}p_2 + \mathbf{ik}p_3 + \mathbf{ji}p_1 + \mathbf{jk}p_3 + \mathbf{ki}p_1 + \mathbf{kj}p_2)$$
$$= 2\mathbf{ij}(\mathbf{k}(p_2 - p_1) + \mathbf{j}(p_1 - p_3) + \mathbf{i}(p_3 - p_2))$$
$$= 2\mathbf{ij}\,\mathbf{1}\times\mathbf{p} . \tag{3}$$

In more conventional terms,
$$[\hat{\sigma}, \mathcal{H}] = 2i\mathbf{ki}(\mathbf{k}(p_2 - p_1) + \mathbf{j}(p_1 - p_3) + \mathbf{i}(p_3 - p_2))$$
$$= 2i\mathbf{k}\,\gamma\times\mathbf{p}$$
$$= 2\gamma_0\,\gamma\times\mathbf{p} . \tag{4}$$

Simultaneously, if **L** is the orbital angular momentum $\mathbf{r}\times\mathbf{p}$,
$$[\mathbf{L}, \mathcal{H}] = [\mathbf{r}\times\mathbf{p}, i\gamma_0\gamma\cdot\mathbf{p} + \gamma_0 m]$$
$$= [\mathbf{r}\times\mathbf{p}, i\gamma_0\gamma\cdot\mathbf{p}] .$$

Taking out common factors,

$$[\mathbf{L}, \mathcal{H}] = i\gamma_0 \, [\mathbf{r}, \gamma.\mathbf{p}] \times \mathbf{p}$$
$$= -k i \, [\mathbf{r}, \mathbf{1}.\mathbf{p}] \times \mathbf{p}$$
$$= -j \, [\mathbf{r}, \mathbf{1}.\mathbf{p}] \times \mathbf{p} .$$

Now,

$$[\mathbf{r}, \mathbf{1}.\mathbf{p}] \, \psi = - i\mathbf{i} \left( x \frac{\partial \psi}{\partial x} - \frac{\partial (x\psi)}{\partial x} \right) - i\mathbf{j} \left( y \frac{\partial \psi}{\partial y} - \frac{\partial (y\psi)}{\partial y} \right) - i\mathbf{k} \left( z \frac{\partial \psi}{\partial z} - \frac{\partial (z\psi)}{\partial z} \right)$$
$$= i\mathbf{1} \, \psi .$$

Hence,

$$[\mathbf{L}, \mathcal{H}] = - i j \, \mathbf{1} \times \mathbf{p} . \tag{5}$$

This, again, can be converted into conventional terms:

$$[\mathbf{L}, \mathcal{H}] = i k i \, \mathbf{1} \times \mathbf{p}$$

$$= i \, \gamma_0 \, \gamma \times \mathbf{p} . \tag{6}$$

Using either (3) and (5) or (4) and (6), we may write

$$[\mathbf{L} - \mathbf{1} / 2, \mathcal{H}] = 0$$

or

$$[\mathbf{L} + \hat{\sigma} / 2, \mathcal{H}] = 0 .$$

Hence, $(\mathbf{L} - \mathbf{1} / 2)$ or $(\mathbf{L} + \hat{\sigma} / 2)$ is a constant of the motion. The physical origin of spin states will be discussed below.

## 12 Helicity

The term

$$\hat{\sigma}.\mathbf{p} = - p_1 - p_2 - p_3 = - p$$

is defined as helicity, and, since it has no vector or quaternion terms, and has only terms of the form $\partial / \partial x$, $\partial / \partial y$, and $\partial / \partial z$ in common with

$$i\gamma_0 \gamma.\mathbf{p} = -j \, (\mathbf{i} p_1 + \mathbf{j} p_2 + \mathbf{k} p_3)$$

and also clearly commutes with $\gamma_0 m = i k m$, then

$$[\hat{\sigma}.\mathbf{p}, \mathcal{H}] = 0$$

and the helicity is a constant of the motion.

For a particle with zero mass, the term $\mathbf{k} \, E + i\mathbf{i} \, p + i\mathbf{j} \, m$ reduces to $\mathbf{k} \, E + i\mathbf{i} \, p$, where $p$ actually represents $\hat{\sigma}.\mathbf{p}$. $E$ also becomes equal to $\pm \, p$. For positive energy states,

$$E = \hat{\sigma}.\mathbf{p} .$$

So the spin is aligned antiparallel to the momentum (has left-handed helicity). Then,

$$ij \, (k \, E + ii \, p) = ij \, (k - ii) \, E = (ii - k) \, E$$

and the spinor wavefunction follows the rule:

$$ij \, u_L = - u_L \, . \qquad (7)$$

For negative energy states,

$$E = - \hat{\sigma} \cdot \mathbf{p} \, ,$$

In this case, the spin is aligned parallel to the momentum (has right-handed helicity). Then,

$$ij \, (k \, E + ii \, p) = ij \, (k + ii) \, E = (ii + k) \, E$$

and the spinor wavefunction follows the rule:

$$ij \, u_R = u_R \, . \qquad (8)$$

From (7) and (8), we may derive the relations

$$\left(\frac{1 - ij}{2}\right) u_L = \left(\frac{1 - \gamma_5}{2}\right) u_R = u_L$$

and

$$\left(\frac{1 - ij}{2}\right) u_R = \left(\frac{1 - \gamma_5}{2}\right) u_R = 0 \, .$$

If we define the right- and left-handed components of the wavefunction, $\psi_R$ and $\psi_L$ by the expressions

$$\psi_R = \left(\frac{1 + \gamma_5}{2}\right) \psi = \left(\frac{1 + ij}{2}\right) \psi$$

and

$$\psi_L = \left(\frac{1 - \gamma_5}{2}\right) \psi = \left(\frac{1 - ij}{2}\right) \psi \, ,$$

we may derive the explicit expressions

$$\left(\frac{1 + ij}{2}\right)(k \, E + ii \, p + ij \, m) = \left(\frac{1 + ij}{2}\right)(k \, (E + p) + m) \qquad (9)$$

and

$$\left(\frac{1 - ij}{2}\right)(k \, E + ii \, p + ij \, m) = \left(\frac{1 - ij}{2}\right)\left(\frac{-km}{E + p}\right)(k \, (E + p) + m) \, . \qquad (10)$$

It is apparent from (9) that $\psi_R = 0$ when $m = 0$ and $E = -p = \hat{\sigma} \cdot \mathbf{p}$.

## 13 The Schrödinger approximation

The derivation of the Schrödinger approximation is purely conventional, and doesn't require the specific use of the nilpotent wavefunction. We take the Dirac equation in the form

$$(i\gamma.\mathbf{p} + m - \beta E)\,\psi = 0$$

and choose, again without loss of generality, the momentum direction $i p_x = \mathbf{p}$. Here again, also, $E$ and $\mathbf{p}$ represent the quantum differential operators, rather than their eigenvalues. This time, we make the conventional choices for $\beta$:

$$\begin{pmatrix} 1 & 0 & 0 & 0 \\ 0 & 1 & 0 & 0 \\ 0 & 0 & -1 & 0 \\ 0 & 0 & 0 & -1 \end{pmatrix}$$

and for $\gamma^1$:

$$\begin{pmatrix} 0 & 0 & 0 & -i \\ 0 & 0 & -i & 0 \\ 0 & i & 0 & 0 \\ i & 0 & 0 & 0 \end{pmatrix}$$

leading to the representation:

$$\begin{pmatrix} E-m & 0 & 0 & -p \\ 0 & E-m & -p & 0 \\ 0 & p & -E-m & 0 \\ p & 0 & 0 & -E-m \end{pmatrix} \begin{pmatrix} \psi_1 \\ \psi_2 \\ \psi_3 \\ \psi_4 \end{pmatrix} = 0.$$

This can be reduced to the coupled equations:

$$(E-m)\phi = p\chi, \qquad (11)$$

and

$$(E+m)\chi = p\phi, \qquad (12)$$

where the bispinors are given by

$$\phi = \begin{pmatrix} \psi_1 \\ \psi_2 \end{pmatrix},$$

and

$$\chi = \begin{pmatrix} \psi_3 \\ \psi_4 \end{pmatrix}.$$

Then, assuming the non-relativistic approximation $E \approx m$, for low $\mathbf{p}$, we obtain

$$\chi \approx \frac{p}{2m} \phi$$

from (12), and

$$(E - m)\phi = \frac{p^2}{2m} \phi, \qquad (13)$$

by substituting into (11). Using the same approximation, $\phi$, here, also becomes $\psi$. Conventionally, of course, the Schrödinger equation excludes the mass energy $m$ from the total energy term $E$, and, in the presence of a potential energy $V$, (13) becomes modified to

$$(E - V)\psi = \frac{p^2}{2m} \psi.$$

Significant, here, is the origin of the factor 2 in the process which square roots the expression $E^2 - m^2$. The origin of the same factor in the derivation of spin from the Dirac equation, is seen in the behaviour of the anticommuting terms which result from this process. In fact, the two factors have precisely the same origin, for we may show, following earlier treatments,[21] that the Schrödinger equation can be used to derive the anomalous magnetic moment of the electron in the presence of a magnetic field **B**. Spin is purely a property of the multivariate nature of the **p** term, and has nothing to do with whether the equation used is relativistic or not. In our operator notation, the Schrödinger equation, whether field-free or in the presence of a field with vector potential **A**, can be written in the form,

$$2mE\psi = \mathbf{p}^2 \psi$$

Using a multivariate, $\mathbf{p} = -i\nabla + e\mathbf{A}$, we derive:

$$2mE\psi = (-i\nabla + e\mathbf{A})(-i\nabla + e\mathbf{A})\psi$$

$$= (-i\nabla + e\mathbf{A})(-i\nabla\psi + e\mathbf{A}\psi)$$

$$= -\nabla^2\psi - ie(\nabla.\psi\mathbf{A} + i\nabla\psi \times \mathbf{A} + \mathbf{A}.\nabla\psi + i\mathbf{A}\times\nabla\psi) + e^2\mathbf{A}^2\psi$$

$$= -\nabla^2\psi - ie(\nabla.\psi\mathbf{A} + 2\mathbf{A}.\nabla\psi + i\psi\nabla\times\mathbf{A}) + e^2\mathbf{A}^2\psi$$

$$= -\nabla^2\psi - ie(\psi\nabla.\mathbf{A} + 2\mathbf{A}.\nabla\psi) + e^2\mathbf{A}^2\psi + e\mathbf{B}\psi$$

$$= (-i\nabla + e\mathbf{A}).(-i\nabla + e\mathbf{A})\psi + e\mathbf{B}\psi$$

$$= (-i\nabla + e\mathbf{A}).(-i\nabla + e\mathbf{A})\psi + 2m\,\mu.\mathbf{B}$$

This is the conventional form of the Schrödinger equation in a magnetic field for spin up. The wavefunction can be either scalar or nilpotent. Reversing the (relative) sign of $e\mathbf{A}$ for spin down, we obtain

$$2mE\psi = (-i\nabla - e\mathbf{A})(-i\nabla - e\mathbf{A})\psi$$

$$= (-i\nabla - e\mathbf{A})(-i\nabla - e\mathbf{A})\psi - 2m\,\mu.\mathbf{B}\,.$$

It is significant that the standard derivation of the Schrödinger equation begins with the classical expression for kinetic energy $p^2/2m$, and that the factor 2 in this equation ultimately carries over into the same factor in the spin term for the electron. It is precisely because the Schrödinger equation is derived via a kinetic energy term that this factor enters into the expression for the spin, and this process is essentially the same as the process which, through the anticommuting quantities of the Dirac equation, makes $(\mathbf{L} + \hat{\sigma}/2)$ a constant of the motion. (Anticommuting operators also introduce the factor 2 in the Heisenberg uncertainty relation for the same reason, the Heisenberg term relating directly to the zero-point energy derived from the kinetic energy of the harmonic oscillator.) In fact, the origin of the factor 2, in all significant cases – classical, quantum, relativistic – is in the virial relation between kinetic and potential energies.[22] In principle, the kinetic energy relation is used when we consider a particle as an object in itself, described by a rest mass $m_0$, undergoing a continuous change. The potential energy relation is used when we consider a particle within its 'environment', with 'relativistic mass', in an equilibrium state requiring a discrete transition for any change. We can consider the kinetic energy relation to be concerned with the action side of Newton's third law, while the potential energy relation concerns both action and reaction. Because of the necessary relation between them, each of these approaches is a proper and complete expression of the conservation of energy. This fundamental relation, as we will show in sections 16 and 17, leads to the significant fact that the nilpotent wavefunctions, in principle, produce a kind of supersymmetry, with the supersymmetric partners not being so much realisable particles, as the couplings of the fermions and bosons to vacuum states.

## 14 The hydrogen atom

The nilpotent version of the Dirac equation has no problem in reproducing such results of standard Dirac theory as, for example, the hyperfine splitting of the energy levels in the hydrogen atom, which can be derived in straightforward terms by the usual textbook methods.[23] Assuming a potential energy of the form $V(r)$, we can write:

$$(kH' + ijm)\psi = (kV(r) + ii\sigma.\mathbf{p} + ijm)\psi\,.$$

Representing $\begin{pmatrix}\psi_1\\\psi_2\end{pmatrix}$ by $\psi'$ and $\begin{pmatrix}\psi_3\\\psi_4\end{pmatrix}$ by $\psi''$ this equation decouples into two parts:

$$(kH' + ijm)\psi' = (kV + ijm)\psi' + ii\sigma.\mathbf{p}\,\psi'' \quad (14)$$
$$(kH' + ijm)\psi'' = ii\sigma.\mathbf{p}\,\psi' + (kV - ijm)\psi'' \quad (15)$$

Collecting terms in (15) and dividing by $2ijm$, we obtain

$$(kH' - kV + 2ijm)\psi'' = ii\sigma.\mathbf{p}\,\psi'$$

$$\left(1 + \frac{kH' - kV}{2ijm}\right)\psi'' = \frac{ii\sigma.\mathbf{p}}{2ijm}\psi'\,.$$

The second term in the left-hand bracket is small, so. assuming the $\psi$'s to be eigenfunctions of the operator $H'$ with energy eigenvalues of $E_n$,

$$\psi'' = \frac{ii\sigma.\mathbf{p}}{2ijm}\left(1 - \frac{kE_n - kV}{2ijm}\right)ii\sigma.\mathbf{p}\,\psi \ .$$

Substituting in (14), we obtain

$$kH'\psi = \left(kV + \frac{ii\sigma.\mathbf{p}}{2ijm}\left(1 - \frac{kE_n - kV}{2ijm}\right)ii\sigma.\mathbf{p}\right)\psi \ . \tag{16}$$

Since

$$\sigma.\mathbf{p} = -\mathbf{i}\sigma_x\frac{\partial}{\partial x} - \mathbf{j}\sigma_y\frac{\partial}{\partial y} - \mathbf{k}\sigma_z\frac{\partial}{\partial z} \ ,$$

$$(ii\sigma.\mathbf{p})(ii\sigma.\mathbf{p}) = p^2$$

and

$$(ii\sigma.\mathbf{p})\,V(r)(ii\sigma.\mathbf{p}) = Vp^2 - \frac{i}{r}\frac{\partial V}{\partial r}(\sigma.\mathbf{r})(\sigma.\mathbf{p})$$

$$= Vp^2 - \frac{i}{r}\frac{\partial V}{\partial r}(\mathbf{r}.\mathbf{p}) - \frac{i}{r}\frac{\partial V}{\partial r}(\sigma.\mathbf{r} \times \mathbf{p}) \ .$$

Equation (16) now becomes

$$kH'\psi = \left(kV + \frac{p^2}{2ijm}\left(1 - \frac{kE_n - kV}{2ijm}\right) - \frac{i}{4m^2}\frac{k}{r}\frac{\partial V}{\partial r}\mathbf{r}.\mathbf{p} + \frac{1}{4m^2}\frac{k}{r}\frac{\partial V}{\partial r}(\sigma.\mathbf{r} \times \mathbf{p})\right)\psi \ . \tag{17}$$

Using $\mathbf{S} = \sigma/2$ and $\mathbf{L} = \mathbf{r} \times \mathbf{p}$, and taking $E_n - V$ as the nonrelativistic kinetic energy term $p^2/2m$, equation (17) becomes

$$kH'\psi = \left(kV + \frac{p^2}{2ijm}\left(1 - \left(\frac{p}{2m}\right)^2\right) - \frac{k}{4m^2}\frac{\partial V}{\partial r}\frac{\partial}{\partial r} + \frac{1}{2m^2}\frac{k}{r}\frac{\partial V}{\partial r}\mathbf{S}.\mathbf{L}\right)\psi \ . \tag{18}$$

The fourth term in the bracket represents $k$ times the interaction energy between spin and orbital moments. With a Coulomb potential due to an isolated nucleus of charge $Ze$,

$$V(r) = -\frac{Ze^2}{4\pi\varepsilon_o r} \ ,$$

this energy becomes

$$\frac{1}{2m^2}\frac{k}{r}\frac{\partial V}{\partial r}\mathbf{S}.\mathbf{L} = \frac{Ze^2}{8\pi\varepsilon_o m^2}\frac{1}{r^3}\mathbf{S}.\mathbf{L} \ .$$

In the standard theory, using the hydrogenic wavefunctions as derived from the nonrelativistic Schrödinger theory, the mean value of $1/r^3$ is calculated as

$$\int_{\infty}^{0} \psi(1/r^3)\psi^* dV = \frac{Z^3 m_r^3 e^6}{(4\pi\varepsilon_o)^3 n^3 l(l + \frac{1}{2})(l + 1)}$$

where $m_r$ is the reduced mass ($\approx m$), and $l \geq 1$, because the term vanishes when $l = 0$ because **L.S** vanishes. The average interaction energy between **L** and **S** then becomes

$$\frac{Ze^2}{8\pi\varepsilon_o m^2} \frac{Z^3 m_r^3 e^6}{(4\pi\varepsilon_o)^3 n^3 l(l + \frac{1}{2})(l + 1)} \mathbf{S.L} = \frac{Z^4 m e^8}{2(8\pi\varepsilon_o)^4 n^3 l(l + \frac{1}{2})(l + 1)} \mathbf{S.L} \ .$$

With the Rydberg constant for infinite mass given by,

$$R_\infty = \frac{me^4}{4\pi(4\pi\varepsilon_o)^2}$$

and the fine structure constant,

$$\alpha = \frac{e^2}{4\pi\varepsilon_o} ,$$

we obtain

$$\frac{Z^4 m e^8}{2(8\pi\varepsilon_o)^4 n^3 l(l + \frac{1}{2})(l + 1)} \mathbf{S.L} = \frac{2\pi R_\infty \alpha^2 Z^4}{n^3 l(l + \frac{1}{2})(l + 1)} \mathbf{S.L} \ ,$$

the energy equation for the hyperfine splitting.

**15 The origin of the factor 2 in spin states**

We have already suggested that all the important factors of 2 (or ½) in classical physics, relativity and quantum physics, can always be attributed to whether one is using potential or kinetic energies, or using just the action side, or both action and reaction sides, of Newton's third law, and that this is a product of whether one is using continuous or discrete solutions, or changing or fixed ones. Through a series of parallelisms and derivations, we can show that this factor 2 comes from the symmetry between the action of an object and the reaction of its environment (whether material or vacuum), and that, while the object taken on its own would show changing behaviour, leading to an integration resulting in a factor ½ (kinetic energy), a conservative 'system' of object plus environment would show unchanging behaviour, expressible in terms of a potential energy term, which is twice the kinetic energy.

    The deeper aspect to this argument is that it makes sense of the boson / fermion distinction in a fundamental way, as well as supersymmetry, vacuum polarization, pair production, renormalization, spin ½, and so on, because the halving of energy in 'isolating' the fermion from its vacuum 'environment' is the same process as mathematically square-rooting the quantum operator via the Dirac equation. The work of Bell et al.[17] on producing integral spins from half-integral spin electrons using the Berry phase is a particular case of this, and it is through this kind of case that we may extend the principle in the direction of supersymmetry.

    The act of defining a rest mass also defines an isolated object, and one cannot define *kinetic* energy in terms of anything but this rest mass. If, however, we take a *relativistic* mass, we are already incorporating the effects of the environment. The

most obvious instance is that of the photon. The photon has no rest mass, only a relativistic mass; $mc^2$ for a photon behaves exactly like a classical potential energy term, as well as having the exact form of a potential energy for a body of mass $m$ and speed $c$. A particular instance is the application of a material gas analogy to an photon gas in producing radiation pressure $\rho c^2/3$. Action and reaction occurs in this instance because the doubling of the value of the energy term comes from the doubling of the momentum produced by the rebound of the molecules / photons from the walls of the container – a classic two-step process, like the two-way speed of light.

The energy involved in both material and photon gas pressure derivations is clearly a potential energy term (the material gas energy has to be halved to relate the kinetic energy of the molecules to temperature), and its double nature is derived from the two-way process which it involves, which is the same thing as saying that it is Newton's action *and* reaction. The same thing happens with radiation reaction, which produces a 'mysterious' doubling of energy $h\nu / 2$ to $h\nu$ in many cases, and also *zitterbewegung* for the electron, which is interpreted as a switching between two states. Feynman and Wheeler also produce a doubling of the contribution of the retarded wave in electromagnetic theory, at the expense of the advanced wave, by assuming that the vacuum behaves as a perfect absorber and reradiator of radiation. In principle, this seems to be equivalent to assuming a filled vacuum for advanced waves (equivalent to Dirac's filled vacuum for antimatter), and relates to previously stated ideas that continuity of mass-energy in the vacuum is related to the unidirectionality of time.[4,5,6]

In the case of material bodies, an old proof of Newton's of the $mv^2/r$ law for centripetal force, and hence of the formula $mv^2$ for orbital potential energy, had the satellite object being 'reflected' off the circle of the orbit, first in a square formation, and then in a polygon with an increasing number of sides, becoming, in the limiting case, a circle. Here, the momentum-doubling action and reaction produce the potential energy formula, and demonstrate the relation between the conservation laws of linear and angular momentum.

Interestingly, though light in free space has velocity $c$, and, through special relativity, has no rest mass, and therefore no kinetic energy, *as soon as you apply a gravitational field*, the light 'slows down', and, at least *behaves* as though it can be treated as a particle with kinetic energy *in the field*. Immediately, then, you can apply a kinetic energy-type equation, and derive the double gravitational bending (ultimately from the 2 in $mc^2/2$) using the very simple $mc^2/2 = GMm (e - 1)/R$, where $e$ is the eccentricity of the hyperbolic orbit.

The result of all these cases is that kinetic energy variation may be thought of as continuous, but starting from a discrete state; potential energy variation, on the other hand, is a discrete variation, starting from a continuous state. Each creates the opposite in its variation from itself. Kinetic energy and potential energy create each other, in the same way as they are related by a numerical relationship.

The existence of two conservation of energy approaches thus has very profound implications, and appears to arise from a very deep stratum in physics. Kinetic energy may be associated with rest mass, because it cannot be defined without it – one could consider light 'slowing down' in a gravitational field as effectively equivalent to adopting a rest mass, and, of course, photons do acquire 'effective masses' in condensed matter. Potential energy is associated with 'relativistic' mass because the latter is *defined* through a potential energy-type term ($E = mc^2$), light in free space being the extreme case, with no kinetic energy / rest mass, and 100 per cent potential energy / relativistic mass. The description, in addition, seems to fit in with the halving that goes on, for a material particle, when we expand its relativistic mass-energy term ($mc^2$) to find its kinetic energy ($mv^2/2$). Thus, if we take the relativistic energy conservation equation

$$E^2 - p^2 - m^2 = 0 ,$$

we can take regard this as a 'relativistic' mass (potential energy equation of the form $E = mc^2$ (treating at one go the particle interacting with its environment), and proceed to quantise to a Klein-Gordon equation, with integral spin. Alternatively, we can separate out the kinetic energy term using the rest mass $m_0$. From

$$E^2 = m_0^2 c^4 \left(1 - \frac{v^2}{c^2}\right) .$$

We take the square root, and obtain

$$E = m_0 c^2 + \frac{mv^2}{2} + \dots .$$

In this case, we are looking at the 'isolated' system of the electron itself. The Schrödinger equation, of course, arises from this approach, quantising $mv^2/2$ in the form $p^2/2m$. Now, as we have shown, using a multivariate form of the momentum operator, $\mathbf{p} = -i\nabla + e\mathbf{A}$, the Schrödinger equation produces the magnetic moment of the electron, with the required half-integral value of spin, the ½ coming from the term $mv^2/2$ or $p^2/2m$; it is also effectively a limiting approximation to the bispinor form of the Dirac equation. In principle, this means that the spin ½ term that arises from the Dirac equation has nothing to do with the fact that the equation is relativistic, but arises from the fundamentally multivariate nature of its use of the momentum operator (equivalent to the use of Pauli matrices). In fact, it comes from the very act of square rooting the energy equation in the same way as that operation produces $mv^2/2$ in the relativistic expansion. The ½ is, in essence, a statement of the act of square-rooting, which is exactly what happens when we split 0 into two nilpotents; the ½ in the Schrödinger approximation is a manifestation of this which we can trace through the ½ in the relativistic binomial approximation.

**16 The fermion and its 'environment'**

Our Dirac equation

$$\left(ik\frac{\partial}{\partial t} + i\nabla + ijm\right)\psi = 0 ,$$

was, of course, obtained by direct square-rooting of the equation

$$E^2 - p^2 - m^2 = 0 ,$$

and leads to a column vector of four terms:

$$(k E + i i\, \mathbf{p} + i j\, m)\, e^{-i(Et - \mathbf{p} \cdot \mathbf{r})}$$
$$(k E - i i\, \mathbf{p} + i j\, m)\, e^{-i(Et - \mathbf{p} \cdot \mathbf{r})}$$
$$(-k E + i i\, \mathbf{p} + i j\, m)\, e^{-i(Et - \mathbf{p} \cdot \mathbf{r})}$$
$$(-k E - i i\, \mathbf{p} + i j\, m)\, e^{-i(Et - \mathbf{p} \cdot \mathbf{r})} ,$$

representing a single quantum state. We have shown that a spin 1 boson creation operator is of the form

$$(kE + i\mathbf{i}\ \mathbf{p} + i\mathbf{j}\ m)\ (-kE + i\mathbf{i}\ \mathbf{p} + i\mathbf{j}\ m)$$
$$(kE - i\mathbf{i}\ \mathbf{p} + i\mathbf{j}\ m)\ (-kE - i\mathbf{i}\ \mathbf{p} + i\mathbf{j}\ m)$$
$$(-kE + i\mathbf{i}\ \mathbf{p} + i\mathbf{j}\ m)\ (kE + i\mathbf{i}\ \mathbf{p} + i\mathbf{j}\ m)$$
$$(-kE - i\mathbf{i}\ \mathbf{p} + i\mathbf{j}\ m)\ (kE - i\mathbf{i}\ \mathbf{p} + i\mathbf{j}\ m)$$

while a spin 0 boson can be represented by

$$(kE + i\mathbf{i}\ \mathbf{p} + i\mathbf{j}\ m)\ (-kE - i\mathbf{i}\ \mathbf{p} + i\mathbf{j}\ m)$$
$$(kE - i\mathbf{i}\ \mathbf{p} + i\mathbf{j}\ m)\ (-kE + i\mathbf{i}\ \mathbf{p} + i\mathbf{j}\ m)$$
$$(-kE + i\mathbf{i}\ \mathbf{p} + i\mathbf{j}\ m)\ (kE - i\mathbf{i}\ \mathbf{p} + i\mathbf{j}\ m)$$
$$(-kE - i\mathbf{i}\ \mathbf{p} + i\mathbf{j}\ m)\ (kE + i\mathbf{i}\ \mathbf{p} + i\mathbf{j}\ m)\ ,$$

each multiplied by the usual exponential form in creating the wavefunction. While the fermion wavefunction is effectively a nilpotent or square root of 0, the boson wavefunction is a product of two nilpotents, each not nilpotent to the other. The multiplications here are scalar multiplications of a 4-component bra vector (composed of the left-hand brackets), representing the particle states, and a ket vector (composed of the right-hand brackets), representing the antiparticle states.

Though we generally consider fermions to follow the Dirac equation, derived, as we have seen, from kinetic energy, and bosons to follow the Klein-Gordon equation, derived ultimately from potential energy, when we look at real situations something remarkable happens. Fermions with spin ½ become spin 1 particles when taken in conjunction with their 'environment' – the Jahn-Teller effect (via the Berry phase) and Aharanov-Bohm effect are examples. A considerable number of phenomena are already associated with the Berry phase, and there are no doubt a large number of others waiting to be discovered.

Bosons with spin 0 or 1 become spin ½ or 3/2 by the reverse process – this is what happens when the gluons interact with quark-gluon plasma to deliver spin ½ or 3/2 to a baryon. It is almost certainly a universal principle that fermions / bosons always produce a 'reaction' within their environment, which couples them to the appropriate wavefunction-changing term, so that the potential / kinetic energy relation can be maintained at the same time as its opposite. We can relate this to the whole process of renormalization, which produces an infinite chain of such couplings through the vacuum. (The related Feynman-Wheeler process of vacuum absorption of radiation, of course, also reduces the infinite electron self-energy to a finite mass.) It is also of significance here that the vacuum wavefunctions we have previously defined are of the complementary forms: $(-kE + i\mathbf{i}\ \mathbf{p} + i\mathbf{j}\ m)$ for fermions, and $(kE + i\mathbf{i}\ \mathbf{p} + i\mathbf{j}\ m)$ for antifermions.

**17 Supersymmetry and renormalisation**

We have seen that ½ spin is always due to kinetic energy (from continuous variation) and we can consider it as the isolated fermion; the unit spin comes from potential energy (stable state) and represents either a boson with two nilpotents (which are not nilpotent to each other) or the equivalent, which is a fermion interacting with its environment – and manifesting Berry phase, radiation reaction, relativistic correction,

Thomas precession, or whatever else is needed to produce a 'conjugate' spin state in its 'environment' (vacuum). It is a question of whether we are treating the action half of Newton's third law, or the action and reaction pair. The 'supersymmetric' partner seemingly comes from this choice of fermion or fermion plus environment approach.

The nilpotent operators defined for fermions are also supersymmetry operators, which produce the supersymmetric partner in the particle itself. The advantage of the supersymmetry concept is that boson contributions and fermion contributions are of opposite sign (with the operators having opposite signs of $E$) and automatically cancel. The particular advantage of this realisation is that there are no extra supersymmetric partners yet to be discovered.

The Jahn-Teller effect throws up the Berry Phase phenomenon because it is treated semi-classically by making an adiabatic approximation (AA). Assuming the time scale for nuclear motions is very much greater than the time scale for electronic transitions, we get a total wave function made up of three factors:
1) A phase factor
2) A factor associated with the electronic coordinates ($q$)
3) A factor associated with the nuclear coordinates ($Q$)

The Jahn-Teller effect couples the $\{Q\}$ and $\{q\}$ motions so that different parts of the total wave function change sign in a coordinated manner to preserve the single-valuedness of the total wave function. Neither the nuclear nor the electronic wavefunction are single-valued by themselves, though the total wavefunction is.

Thus, a sign change occurs when a fermion rotates through $2\pi$, but if the relationship between fermion and 'the rest of the universe' is like the total wave function of the Jahn-Teller effect then there is a preservation of single-valuedness. The Jahn-Teller effect is also the spontaneous breaking of a point-group symmetry, the electronic part of the Hamiltonian producing a Mexican hat potential upon which the nuclei move. This adiabatic potential is the basis of the symmetry breaking. It is the Jahn-Teller effect that couples the nuclear coordinates to the electronic coordinates, and it may be that the same thing be going on in continuous symmetries to couple electrons to the vacuum. (Certain vibronic – vibrational / electronic – interactions take up continuous group symmetries to first order in the Jahn-Teller effect, and this may have parallels with the Higgs mechanism.)

If this is true, the coupling of the vacuum to electrons could be generating 'boson-images' and vice versa. Presumably the required loop diagrams that lead to renormalisation (cancellation of fermion with boson loops) would then have an explanation without us needing to postulate the existence of extra boson/fermion equivalents (photon/photino, etc.). 'Supersymmetry' may be part of a much more general pattern. Bosons and fermions seem to require 'partner states' as much as potential and kinetic energy are needed to fully describe conservation.

The $Q$ generator for supersymmetry is simply the term ($k E + i i \mathbf{p} + i j m$). To convert bosons to fermions, or antifermions to bosons, we multiply by ($k E + i i \mathbf{p} + i j m$) (the **p** could, of course, be + or –). To convert bosons to antifermions, or fermions to bosons, we multiply by ($-k E + i i \mathbf{p} + i j m$). These are $Q$ and $Q\dagger$, and (taking the normalisation that we have previously introduced for the vacuum operator) the anticommutator becomes effectively $E$, the Hamiltonian. Of course, if we write out the operators in full, they will be respectively four-term bra and ket vectors, with the $E$ and **p** values going through the usual cycle of + and – values.

Using the 'environment' principle, we can imagine an infinite series of interacting terms of the form ($k E + i i \mathbf{p} + i j m$), ($k E + i i \mathbf{p} + i j m$) ($-k E + i i \mathbf{p} + i j m$), ($k E + i k \mathbf{p} + i j m$) ($-k E + i i \mathbf{p} + i j m$)( $k E + i i \mathbf{p} + i j m$), ($k E + i i \mathbf{p} + i j m$) ($-k E + i i \mathbf{p} + i j m$)( $k E$

+ *ii***p** + *ij*m) (–*kE* + *ii***p** + *ij*m), etc., where selection of the appropriate terms will lead to a cancellation of the boson and fermion loops of opposite sign. The (*kE* + *ii***p** + *ij*m) and (–*kE* + *ii***p** + *ij*m) or bra and ket factors are an expression of the behaviour of the vacuum. To convert bosons to fermions we multiply by (*kE* + *ii***p** + *ij*m) (the **p** could, of course, be + or –). To convert bosons to antifermions, or fermions to bosons, we multiply by (–*kE* + *ii***p** + *ij*m).

In effect, the vacuum state must act like a 'mirror image' to the fermion.

$$a \,|\, 0 \rangle = (kE + ii\mathbf{p} + ijm)\, k\, (kE + ii\mathbf{p} + ijm)$$

not only looks like a 'mirror image' when you write it down, but is the expression of part of an infinite regression of images of the form

$$(kE + ii\mathbf{p} + ijm)\, k\, (kE + ii\mathbf{p} + ijm)\, k\, (kE + ii\mathbf{p} + ijm)\, k\, (kE + ii\mathbf{p} + ijm)\, \ldots$$

Though it is possible to construct a vacuum state that these operators act upon, one gets a vacuum state that is dependent on the operator that is acting on it. Thus, the vacuum state of the state (*kE*$_1$ + *ii***p**$_1$ + *ij*m) turns out to be *k* (*kE*$_1$ + *ii***p**$_1$ + *ij*m). In addition, however,

$$(kE + ii\mathbf{p} + ijm)\, k\, (kE + ii\mathbf{p} + ijm)\, k\, (kE + ii\mathbf{p} + ijm)\, k\, (kE + ii\mathbf{p} + ijm)\, \ldots$$

is the same as

$$(kE + ii\mathbf{p} + ijm)\, (-kE + ii\mathbf{p} + ijm)\, (kE + ii\mathbf{p} + ijm)\, (-kE + ii\mathbf{p} + ijm)\, \ldots$$

The infinite series of creation acts on vacuum is the same as the mechanism for fermion / boson / fermion / boson ... creation (supersymmetry, renormalization), and is only true if the series is infinite, because each bracket has to be postmultiplied by *k* to make it work. The spin terms **p** must, of course, be of the same sign to produce spin 1 bosons – spin 0, as in the case of the mass-generating Higgs, would break the sequence. The 'reflection' process also implies an infinite range of virtual *E* values in vacuum (though, presumably, only of one sign in the ground state).

Here, then, we may even have an explanation of the mechanism of renormalisation. It is also worth noting that the Klein-Gordon equation automatically applies to fermions (because it simply involves pre-multiplication of zero by a nilpotent differential operator), while the Dirac equation applies to 'spin 1' particles (in the particle + environment sense – like a method of images), leading to the Aharonov-Bohm effect, the Jahn-Teller effect, *zitterbewegung*, etc. With the 'environment' principle we don't need to suppose we have an *extra* set of bosons or fermions – the coupling to the vacuum is automatic in a series of entangled states.

In splitting away a fermion from the 'system' we introduce a coupling as a mathematical description of being able to split it away; the coupling is to the rest of the universe to make the total wavefunction single-valued. Fermions by themselves cannot have single-valued wavefunctions, if this *total* wavefunction is to be single-valued. It may be possible, as we have supposed, to *predict* effects of the Jahn-Teller, Aharonov-Bohm type, where the fermion acquires a single-valued wavefunction by interaction with its 'environment'. There must be other effects of this type which can be predicted and detected – maybe many others.

The reverse effect must also be possible, of bosons coupling to an 'environment' to produce fermion-like states. The most immediately likely possibility is the behaviour of gluons in producing the baryon total spin of ½ or 3/2. The six-component baryon wavefunction has something equivalent to ($kE \pm i\mathbf{i}p_x + \mathbf{i}\mathbf{j}m$) ($kE \pm i\mathbf{i}p_y + \mathbf{i}\mathbf{j}m$) ($kE \pm i\mathbf{i}p_z + \mathbf{i}\mathbf{j}m$), and the $p_x$, $p_y$, $p_z$ and $\pm$ represent the six degrees of freedom for **p**. We can imagine the **p** rotating through the three spatial positions leaving terms like ($kE \pm i\mathbf{i}p + \mathbf{i}\mathbf{j}m$) ($kE + \mathbf{i}\mathbf{j}m$) ($kE + \mathbf{i}\mathbf{j}m$); ($kE + \mathbf{i}\mathbf{j}m$) ($kE \pm i\mathbf{i}p + \mathbf{i}\mathbf{j}m$) ($kE + \mathbf{i}\mathbf{j}m$); etc., which effectively produce something like a fermion combined with Bose-Einstein condensate. This is correct, but how do the *gluons* do it? Presumably, in effect, by the gluon 'transferring' the **p** between one ($kE + \mathbf{i}\mathbf{j}m$) and another – to use crudely mechanistic terms – becoming a boson of spin 1 with an effective contribution from the 'environment' due to the gluon sea making it transfer spin ½.

Another aspect of the process is that *dimensionality*, in general, introduces two orders of meaning in a parameter – of the value (as in length/time or charge/mass), and of the squared value (as in Pythagorean/vector addition of space dimensions, or space and time, or energy and momentum, or charges/masses 'interacting' to produce forces). In a sense we are doing this with fermion and boson wavefunctions, one type being a 'square root' of the other.

**18 The creation of the Dirac state**

We have now reached the point where we can suggest how the algebra is involved in the creation of the Dirac state. We have four fundamental parameters, time, space, mass, charge, which require the respective algebraic units *i*, **i**, **j**, **k**, 1, *i*, *j*, *k*. However, the fundamental units of the *algebra* which they entail is created from an anticommuting pentad, with specific rules concerning the two 3-dimensional arrangements, **i**, **j**, **k**, *i*, *j*, *k*.[9] To reduce the 8 pure units to 5 composite ones, we need to map one of them onto the other three *parameters*. So, we map charge onto space, time and mass, *w*, *s*, *e*, each taking one of the parameters (though we could no doubt also obtain something useful by mapping space onto time, charge and mass). This gives us an algebra for the composites of the form *k*i, *i***i**, *i***j**, *i***i**, *j*, which is, of course, the Dirac algebra (though we usually there multiply by an extra *i* purely for convenience).

But what have we done physically? We have, by putting charge components onto time, space and mass, actually introduced quantization (for charge is, of course, itself quantized), and since the charges are conserved quantities, we have also created a *quantum state* with fixed *E*-**p**-*m*. The system, of course, is already relativistic because our original units were, and the exponential term serves to preserve the required variation in time and space. *E*-**p**-*m* is the direct result of mapping conserved unit charges onto time, space and mass. The concept of rest mass is only created by the act of 'quantization'.

Charge, being irrotational, because of the separate conservations of the weak, strong and electric charges, should not rotate the values of *w*-*s*-*e*, but the actual mapping is arbitrary. Hence we have to have rules for 'charge accommodation'. The *k*, *i*, *j* in the Dirac equation *are*, in principle, the charges (and this, in a sense, justifies the Dirac equation's otherwise arbitrary interpretation of –*E* states as antiparticles), and the *E*-**p**-*m* Dirac state is actually *created* by *w*-*s*-*e*. This is a direct link between *E*-**p**-*m* and *w*-*s*-*e*, in addition to justifying the fundamental nature of the time, space, mass, charge paradigm, and explaining in physical terms why we have such a thing as a Dirac state.

*E*-**p**-*m* and *w*-*s*-*e* reflect different ways of combining the 8 parts to make 5 (we may refer to this process as 'compactification' or 'folding'), and so one might expect different mathematical structures as we make the combination *E*-**p**-*m*-like or *w*-*s*-*e*-like. Doing the 'compactification' in two different ways doesn't mean that the resulting composite notions will map directly onto each other. *E* and **p** are fixed in a Dirac state, but they have an infinite range of possible values. This is not true of *w* and *s*, and we cannot necessarily infer that the value of *E*/**p** will reflect that of *w*/*s* (or even its presence). *E*-**p**-*m* is certainly easier to generate than *w*-*s*-*e*, but one can see that the process of 'folding' the charge quaternions onto time, space, mass can be said to make one charge time-like, another space-like, and another mass-like. It may be, however, that the 'folding' process may be of the charge quaternions onto the Dirac *E*-**p**-*m*, rather than the original time, space, mass.

*E*-**p**-*m* is discrete, being made so in the process of 'compactifying' the charge. If imaginary numbers create a discontinuous $U(5)$,[24] then this is because in *E*-**p**-*m* we have a real quantized mass and an imaginary quantized *E*. Imaginary *i* is not only necessary to complete the Dirac pentad, it is also necessary to fermions because *E* / *w* is necessary to fermions. It is also interesting that the group is $U(5)$ because the gravity operator ('mass') has been automatically included in the process of 'compactification'. Imaginary *i* must introduce discreteness in the square-rooting process which produces *iE*, and this links up with the fermion / boson distinction – square-rooting the energy equation introduces discreteness (the fermion, with rest mass) in a continuous variation (the kinetic energy).